\documentstyle[12pt]{article}

\catcode`\@=11
\def\@citex[#1]#2{%
\if@filesw \immediate \write \@auxout {\string \citation {#2}}\fi
\@tempcntb\m@ne \let\@h@ld\relax \def\@citea{}%
\@cite{%
  \@for \@citeb:=#2\do {%
    \@ifundefined {b@\@citeb}%
      {\@h@ld\@citea\@tempcntb\m@ne{\bf ?}%
      \@warning {Citation `\@citeb ' on page \thepage \space undefined}}%
      {\@tempcnta\@tempcntb \advance\@tempcnta\@ne%
      \@tempcntb\number\csname b@\@citeb \endcsname \relax%
      \ifnum\@tempcnta=\@tempcntb %   Number follows previous--hold on to it
        \ifx\@h@ld\relax%
          \edef \@h@ld{\@citea\csname b@\@citeb\endcsname}%
        \else%
          \edef\@h@ld{\ifmmode{-}\else--\fi\csname b@\@citeb\endcsname}%
        \fi%
      \else%   %  non-successor--dump what's held and do this one
        \@h@ld\@citea\csname b@\@citeb \endcsname%
        \let\@h@ld\relax%
      \fi}%
    \def\@citea{,\penalty\@highpenalty\,}%
  }\@h@ld
}{#1}}

\def\@citeb#1#2{{[#1]\if@tempswa , #2\fi}}
\def\@citeu#1#2{{$^{#1}$\if@tempswa , #2\fi }}
\def\@citep#1#2{{#1\if@tempswa , #2\fi}}

\def\bcites{         % cite with []'s
	\catcode`\@=11
	\let\@cite=\@citeb
	\catcode`\@=12
}

\def\upcites{         % cite with exponents
	\catcode`\@=11
	\let\@cite=\@citeu
	\catcode`\@=12
}

\def\plaincites{      % cite without brackets
	\catcode`\@=11
	\let\@cite=\@citep
	\catcode`\@=12
}

\newcount\hour
\newcount\minute
\newtoks\amorpm
\hour=\time\divide\hour by 60
\minute=\time{\multiply\hour by 60 \global\advance\minute by-\hour}
\edef\standardtime{{\ifnum\hour<12 \global\amorpm={am}%
        \else\global\amorpm={pm}\advance\hour by-12 \fi
        \ifnum\hour=0 \hour=12 \fi
        \number\hour:\ifnum\minute<10 0\fi\number\minute\the\amorpm}}
\edef\militarytime{\number\hour:\ifnum\minute<10 0\fi\number\minute}

\def\draftlabel#1{{\@bsphack\if@filesw {\let\thepage\relax
   \xdef\@gtempa{\write\@auxout{\string
      \newlabel{#1}{{\@currentlabel}{\thepage}}}}}\@gtempa
   \if@nobreak \ifvmode\nobreak\fi\fi\fi\@esphack}
        \gdef\@eqnlabel{#1}}
\def\@eqnlabel{}
\def\@vacuum{}
\def\marginnote#1{}
\def\draftmarginnote#1{\marginpar{\raggedright\scriptsize\tt#1}}
\def\draft{
	\pagestyle{plain}
	\overfullrule=2pt
        \oddsidemargin -.5truein
        \def\@oddhead{\sl \phantom{\today\quad\militarytime} \hfil
        \smash{\Large\sl DRAFT} \hfil \today\quad\militarytime}
        \let\@evenhead\@oddhead
        \let\label=\draftlabel
        \let\marginnote=\draftmarginnote
        \def\ps@empty{\let\@mkboth\@gobbletwo
        \def\@oddfoot{\hfil \smash{\Large\sl DRAFT} \hfil}
        \let\@evenfoot\@oddhead}
        \def\@eqnnum{(\theequation)\rlap{\kern\marginparsep\tt\@eqnlabel}%
        \global\let\@eqnlabel\@vacuum}  }

\def\blackfonts{
        \font\blackboard=msbm10 scaled\magstep1
        \font\blackboards=msbm8
        \font\blackboardss=msbm6
}

\def\nblack{            % For people without blackboard fonts
	\def\ZZ{{Z \n{10} Z}}
	\def\NN{{N \n{14} N}}
	\def\CC{{C \n{11} C}}
	\def\RR{{R \n{11} R}}
	\def\QQ{{Q \n{12} Q}}
	\def\PP{{P \n{11} P}}
}

\def\prep{         % twocolumn.sty  Changed my Marek and Neil
	\catcode`\@=11
        \input art10.sty
	\catcode`\@=12
	
	\let\small\null
	\def\blackfonts{
        	\font\blackboard=msbm10
        	\font\blackboards=msbm7
        	\font\blackboardss=msbm5
	}
	\let\sl\it
        \twocolumn
        \sloppy
        \voffset=-2.54truecm
        \hoffset=-2.54truecm
        \flushbottom
        \parindent 1em
        \leftmargini 2em
        \leftmarginv .5em
        \leftmarginvi .5em
        \marginparwidth 48pt
        \marginparsep 10pt
        \setlength{\columnsep}{2truecm}
        \setlength{\textwidth}{25.4truecm}
        \setlength{\textheight}{17truecm}
        \baselineskip=16pt
        \oddsidemargin .18truein
        \evensidemargin .17truein
}

\def\eqalign#1{\null\,\vcenter{\openup\jot\m@th
  \ialign{\strut\hfil$\displaystyle{##}$&$\displaystyle{{}##}$\hfil
      \crcr#1\crcr}}\,}
\def\eqalignno#1{\displ@y \tabskip\centering
  \halign to\displaywidth{\hfil$\@lign\displaystyle{##}$\tabskip\z@skip
    &$\@lign\displaystyle{{}##}$\hfil\tabskip\centering
    &\llap{$\@lign##$}\tabskip\z@skip\crcr
    #1\crcr}}

\def\section{\@startsection {section}{1}{\z@}{3.ex plus 1ex minus
 .2ex}{2.ex plus .2ex}{\large\bf}}
\def\subsection{\@startsection{subsection}{2}{\z@}{2.75ex plus 1ex minus
 .2ex}{1.5ex plus .2ex}{\bf}}

\def\appendix{{\newpage\section*{Appendices}}\let\appendix\section%
        {\setcounter{section}{0}
        \gdef\thesection{\Alph{section}}}\section}

\def\abstract{\if@twocolumn
\section*{Abstract}
\else %\small
\begin{center}
{\bf Abstract\vspace{-.5em}\vspace{0pt}}
\end{center}
\quotation
\fi}

\catcode`\@=12

\def\ansatz{{\it ansatz\/}}

\def\apriori{{\it a priori\/}}

\def\etal{{\it et al.\/}}

\def\ie{\hbox{\it i.e.\/}}

\def\noj#1,#2,{{\bf #1} (19#2)\ }
\def\jou#1,#2,#3,{{\sl #1\/ }{\bf #2} (19#3)\ }
\def\ann#1,#2,{{\sl Ann.\ Physics\/ }{\bf #1} (19#2)\ }
\def\cmp#1,#2,{{\sl Comm.\ Math.\ Phys.\/ }{\bf #1} (19#2)\ }
\def\cq#1,#2,{{\sl Class.\ Quantum Grav.\/ }{\bf #1} (19#2)\ }
\def\cqg#1,#2,{{\sl Class.\ Quantum Grav.\/ }{\bf #1} (19#2)\ }
\def\ijmp#1,#2,{{\sl Int.\ J.\ Mod.\ Phys.\/ }{\bf A#1} (19#2)\ }
\def\jmp#1,#2,{{\sl J.\ Math.\ Phys.\/ }{\bf #1} (19#2)\ }
\def\grg#1,#2,{{\sl Gen.\ Rel.\ Grav.\/ }{\bf #1} (19#2)\ }
\def\mpl#1,#2,{{\sl Mod.\ Phys.\ Lett.\/ }{\bf A#1} (19#2)\ }
\def\nc#1,#2,{{\sl Nuovo Cim.\/ }{\bf #1} (19#2)\ }
\def\np#1,#2,{{\sl Nucl.\ Phys.\/ }{\bf B#1} (19#2)\ }
\def\pl#1,#2,{{\sl Phys.\ Lett.\/ }{\bf #1B} (19#2)\ }
\def\pla#1,#2,{{\sl Phys.\ Lett.\/ }{\bf #1A} (19#2)\ }
\def\pr#1,#2,{{\sl Phys.\ Rev.\/ }{\bf #1} (19#2)\ }
\def\prd#1,#2,{{\sl Phys.\ Rev.\/ }{\bf D#1} (19#2)\ }
\def\prl#1,#2,{{\sl Phys.\ Rev.\ Lett.\/ }{\bf #1} (19#2)\ }
\def\prp#1,#2,{{\sl Phys.\ Rept.\/ }{\bf #1C} (19#2)\ }
\def\ptp#1,#2,{{\sl Prog.\ Theor.\ Phys.\/ }{\bf #1} (19#2)\ }
\def\ptpsup#1,#2,{{\sl Prog.\ Theor.\ Phys.\/ Suppl.\/ }{\bf #1} (19#2)\ }
\def\rmp#1,#2,{{\sl Rev.\ Mod.\ Phys.\/ }{\bf #1} (19#2)\ }
\def\yadfiz#1,#2,#3[#4,#5]{{\sl Yad.\ Fiz.\/ }{\bf #1} (19#2) #3%
\ [{\sl Sov.\ J.\ Nucl.\ Phys.\/ }{\bf #4} (19#2) #5]}
\def\zh#1,#2,#3[#4,#5]{{\sl Zh.\ Exp.\ Theor.\ Fiz.\/ }{\bf #1} (19#2) #3%
\ [{\sl Sov.\ Phys.\ JETP\/ }{\bf #4} (19#2) #5]}

\def\eq#1{.~(\ref{#1})}
\def\noeq#1{(\ref{#1})}
\def\beq{\begin{equation}}
\def\eeq{\end{equation}}
\def\beqar{\begin{eqnarray}}
\def\eeqar{\end{eqnarray}}

\def\nfrac#1#2{{\displaystyle{\vphantom1\smash{\lower.5ex\hbox{\small$#1$}}%
        \over\vphantom1\smash{\raise.25ex\hbox{\small$#2$}}}}}

\def\d#1{{}_{#1}}
\def\p#1{\mskip#1mu}
\def\n#1{\mskip-#1mu}
\def\stop{\p6.}
\def\comma{\p6,}
\def\semi{\p6;}

\def\eqand{\p8 {\rm and}}

\def\to{\rightarrow}

\def\longlongrightarrow{\relbar\joinrel\relbar\joinrel\rightarrow}

\def\onArrow#1{\mathrel{\mathop{\longlongrightarrow}\limits^{#1}}}

\def\lae{\mathrel{\mathop{\smash{\lower .5 ex \hbox{$\stackrel<\sim$}}}}}
\def\lae{\mathrel{\mathop{\smash{\lower .5 ex \hbox{$\stackrel>\sim$}}}}}

\def\ket#1{\left| #1 \right\rangle}

\def\pa{\partial}

\def\l:{\mathopen{:}\,}
\def\r:{\,\mathclose{:}}
\def\sech{\mathop{\rm sech}\nolimits}

\catcode`\@=11
\def\theequation{\arabic{equation}}
\catcode`\@=12

\nblack
\bcites

\nblack

\catcode`\@=11
\def\theequation{\thesection.\arabic{equation}}
\@addtoreset{equation}{section}
\@addtoreset{footnote}{section}
\@addtoreset{footnote}{subsection}
\catcode`\@=12

\def\sltr{$SL(2,\RR)$}
\def\sltrouo{$SL(2,\RR)/U(1)$}
\def\half{\nfrac12}
\def\thrhalf{\nfrac32}
\def\d{{\rm d}}
\def\tdot{{\dot T}}
\def\fdot{{\dot f}}
\def\calo{{\cal O}}
\def\eipx{e^{\sqrt2 \p1 i \p1 p \p1 x}}
\def\eix#1{e^{\sqrt2 \p1 i \p1 #1 \p1 x}}
\def\ep#1{e^{#1 \sqrt2 \p1 \varphi}}
\def\epa#1#2{e^{#1 \sqrt2 \p1 #2 \p1 \varphi}}
\def\F{F\n4\left(\half, \half, 1, 1 - z\right)}
\def\Fsq{F^2\n4\left(\half, \half, 1, 1 - z\right)}
\def\Fd{F\n4\left(\half, \half, 2, 1 - z\right)}
\def\Fdsq{F^2\n4\left(\half, \half, 2, 1 - z\right)}
\def\Fp{F\n4\left(\half + |p|,\half + |p|,1 + |p|,1-z\right)}

\begin{document}
\begin{titlepage}

\begin{center}
May 3, 1993\hfill    TAUP--2046--93 \\
\hfill    hep-th/9305003

\vskip 1 cm

{\large \bf
The spectrum of the 2D Black Hole\\
}
\vskip .1 cm
{
\bf
or Does the 2D black hole have tachyonic or W--hair?
}

\vskip .5 cm
{
	  Neil Marcus${}^*$ {\it and\/} Yaron Oz\footnote{
Work supported in part by the US-Israel Binational Science Foundation,
and the Israel Academy of Science.  E-Mail:
NEIL@HALO.TAU.AC.IL, YARONOZ@CCSG.TAU.AC.IL}
}
\vskip 0.2cm

{\sl
School of Physics and Astronomy\\Raymond and Beverly Sackler Faculty
of Exact Sciences\\Tel-Aviv University\\Ramat Aviv, Tel-Aviv 69978, ISRAEL.
}

\end{center}

\vskip .5 cm

\begin{abstract}

We solve the equations of motion of the tachyon and the discrete states
in the background of Witten's semiclassical black hole and in the exact
2D dilaton-graviton background of Dijkgraaf \etal{}  We find the exact
solutions for weak fields, leading to conclusions in disagreement with previous
studies of tachyons in the black hole.  Demanding that a state in
the black hole be well behaved at the horizon implies that it must tend
asymptotically to a combination of a Seiberg and an anti-Seiberg $c=1$
state.  For such a state to be well behaved asymptotically, it must
satisfy the condition that neither its Seiberg nor its anti-Seiberg
Liouville momentum is positive.  Thus, although the free-field BRST
cohomologies of the underlying \sltrouo{} theory is the same as that of
a $c=1$ theory, the black hole spectrum is drastically truncated: {\it
There are no $W_\infty$ states, and only tachyons with $x$-momenta
$|p_{\rm tach}| \le |m_{\rm tach}|$ are allowed.\/}  In the Minkowski
case only the static tachyon is allowed.  The black hole is stable to
the back reaction of these remaining tachyons, so they are good
perturbations of the black hole, or ``hair''.  However, this leaves
only 3 tachyonic hairs in the black hole and 7 in the exact solution!
Such sparse hair is clearly irrelevant to the maintenance of coherence
during black hole evaporation.

\end{abstract}
\newpage

\end{titlepage}

\section{Introduction}

Black hole solutions of two-dimensional string theory
\cite{Witten,Wadia,Dijkgraaf} have attracted much attention in the last
couple of years.  Black holes are very interesting since they provide
an arena for studying the interplay between gravity and quantum
theory.  Since string theories can be solved, at least in principle,
understanding black holes in string theories could shed some light
at the quantum level on deep issues such as
singularities,  the coherence of black hole evaporation, no hair
theorems and cosmic censorship Evidently, our real interest is in four
dimensions and the hope is that the two-dimensional case is rich enough
to teach us something about that.

The two-dimensional black holes are constructed using the fact that
the \sltrouo{} gauged WZW theory at level $k=9/4$ is a good background to
the bosonic string \cite{Witten}.  The usual black hole spacetime can
be found from the semi-classical approximation to this coset theory
\cite{Witten}, or by
by solving the $\beta$--function
equations to lowest order in $\alpha'$ \cite{Wadia} in the
sigma-model approach to strings in curved spacetime.
An exact dilaton-graviton background has also been proposed by
Dijkgraaf \etal{} by expressing the $L_0, \bar{L}_0$
Virasoro generators of the \sltrouo{} gauged WZW theory as
differential operators on the \sltr{} group manifold \cite{Dijkgraaf},
and it has been verified that this background solves the
$\beta$--function equation perturbatively to 4 loops \cite{Tseytlin,Liverpool}.
This spacetime also has the
essential properties of a black hole.

Since the black hole is a 2-dimensional solution to the string, it is not
surprising that
there are many relations between it and the $c=1$ theory:
First, the black hole tends asymptotically to a flat space
with a linear dilaton, which is the target space of the $c=1$ theory without
a cosmological constant.  In addition, the $c=1$ theory can be perturbed to
the black hole by the anti-Seiberg ${\overline W}_{1,0}^- W_{1,0}^-$
discrete state.  In fact, using a free-field representation of the current
algebra,  the spectrum of the \sltrouo{} gauged WZW theory has been found to
be in a one-to-one correspondence with the $c=1$ theory \cite{Eguchi}, with
infinitely many tachyon and $W_\infty$ states.
(There is also a calculation directly in the Kac-Moody module which
has extra states \cite{Distler}.)
This correspondence has led to the expectation that the black hole
hole has an infinite amount of ``W--hair'', giving it infinitely many
conserved quantum numbers, and that this W--hair can maintain
quantum coherence during the evaporation of the black hole \cite{Ellis}.

However, here the $c=1$ theory should provide us with a warning:  The
BRST free-field cohomology is not necessarily the same as the spectrum
of the theory.  Thus, when the cosmological constant $\mu$ is turned
on, it is accepted that the spectrum of the $c=1$ theory
is restricted to only the
Seiberg states \cite{Seiberg}, with anti-Seiberg states such as the
black-hole operator destroying the background.  In the black hole, one
might expect the truncation of the spectrum to be much stronger
because of the classical ``no-hair theorems'' of black holes.
By exactly solving the linearized equations of motion of the tachyon
and $W_\infty$ states in the background of the black hole,
we shall see that the black hole actually has
{\it no\/} W--hair, and only very few tachyonic hairs, depending on
exactly which version of the theory is being considered.

We should point out that
perturbing the black hole solution by static
tachyons in
the linearized approximation sigma-model approach has been studied by
various authors\cite{Lykken,Minic,Rama}.
It has been
claimed that the black hole structure is changed due to the
perturbation---that the horizon is split into two yielding a structure
similar to
Reissner-Nordstrom black hole\cite{Lykken,Rama}, and that a curvature
singularity develops at the horizon\cite{Rama}.
What is common to all these works is that the equation
of motion for the tachyon in the dilaton-graviton black hole background
was not completely solved, and that the study of the back reaction was
based on a tachyon configuration that diverges at the horizon.
Since this is an invalid solution, we disagree with their results.
The stability of the black hole against tachyon perturbation was studied
in a somewhat different context in \cite{Elitzur}. We disagree with their
result concerning the change of the radius of the cigar shaped black hole
manifold due to the tachyon perturbation.

The paper is organized as follows: In section~2 we review the basics of
the two dimensional string black hole, making comments and establishing
conventions.  In section~3 we solve the equation describing a static
tachyon in the approximate black hole background.  We discuss the
appropriate boundary conditions to impose asymptotically and at the
horizon of the black hole, and see that the tachyon asymptotically
becomes a mixture of the zero-momentum tachyon and the ``cosmological
constant'' operator $\varphi \ep-$ of the 2D string.  We then study its
back reaction on the dilaton-graviton system, and verify the stability of the
black hole.  All these calculations are performed in the linear-dilaton
gauge, and are repeated in the conformal gauge in the appendix.  In
section~4 we study tachyons with non-zero momenta.
We find that
in the Minkowski black hole only static tachyons can exist, and in the
Euclidean black hole only the 3
tachyons with momenta $|p_{\rm tach}| \le |m_{\rm tach}|$ are well behaved.
We show that
the black hole is again stable to the back-reaction of these tachyons.  In
section~5 we study the massive $W_\infty$ states in the black hole
background, after making an \ansatz{} as to how to derive them from the
asymptotic $c=1$ states.  We solve their equations of motion and see
that no such states can exist in the black hole, so the black hole has
no W--hair.  In section~6 we analyse the tachyon and W--hair in the
exact dilaton-graviton background, finding similar results to those of the
black-hole, except that there are now 7 allowed tachyons with
$|p_{\rm tach}| \le |m_{\rm tach}|$ in the
Euclidean case.  Section~7 is devoted to discussions and conclusions.

\section{Review of the black hole in 2 dimensions}

\subsection{ The sigma-model}

The chiral sector of
$c=1$ string theory describes states with ghost numbers ranging from $-1$ to
$+1$ \cite{Lian,Pilch,Witalone}.  Those with ghost number 0, with which
we shall be concerned, are the tachyons which can
have an arbitrary $x$-momentum, and an infinite number of discrete
states ``$W_\infty$'' states at fixed $x$-momentum \cite{Witalone,Kleb}.
Both the tachyons and the discrete states come in ``Seiberg'' and
``anti-Seiberg'' versions.
Precisely which momenta (and windings) of the tachyon and which
discrete states survive to the full theory depend on the radius of
compactification of the $x$-coordinate of the theory \cite{Mukhi}.

There are two approaches to constructing spacetime field theories for
this---or any other---string theory; string field theory (SFT)
or low-energy
effective actions.
The SFT approach contains the infinite
number of fields of the higher-dimensional strings with a large
gauge-invariance, but has the disadvantage that it describes all these fields
perturbatively.  It is thus hard to see non-trivial geometries in this
language.
The effective action approach to the string is somewhat orthogonal to that of
the SFT:  here one considers the theory as an expansion
in $\alpha'$ (set to 2 in our notation),
where it should be
dominated by the massless fields.  The
massless discrete states are then represented
by a metric $G_{\mu \nu}$ and a dilaton field $\Phi$, which can be
large compared to their flat-space values.  (There is no perturbative
axion field, since we are in two dimensions.)  One can also consider the
tachyonic and massive states in this
framework, but again only as perturbations.  This procedure has been partially
carried for the tachyon, which is
described by a tachyon field $T(x,\varphi)$ as in the SFT.  We shall
postpone a discussion of the discrete states to section~\ref{discrete-sec}.

To lowest order in $\alpha'$, the
sigma-model action of the dilaton-graviton-tachyon system is given by
\cite{Callen,tach}
\beq
S = \int{\rm d}^2x \, e^{-2\Phi}\sqrt{G} \left(
R - 4(\nabla\Phi)^2 + (\nabla T)^2
+ V(T) + c/3 \right) \label{sig} \comma
\eeq
and the equations of motion for $G_{\mu\nu}$, $\Phi$ and $T$
are
\beq
\eqalign{
& R_{\mu\nu} - 2\nabla_{\mu}\nabla_{\nu}\Phi + \nabla_{\mu}T\nabla_{\nu}T
= 0 \comma \cr
& R + 4(\nabla\Phi)^2 -4\nabla^2\Phi +(\nabla T)^2 + V(T) + c/3 = 0 \comma \cr
& {\nabla}^2T -2 \nabla\Phi\nabla T -\nfrac12  \, V'(T) = 0\comma
\label{beteq}
}
\eeq
respectively.
In these formulae,
$c$ is the central charge ($c=D-26= -24$) and $V(T)$ is the
potential of the tachyon.
The potential begins with the mass term $V = -2
T^2$, but the existence of higher-order terms has has been the subject of
some controversy \cite{Banks}.   We shall restrict ourselves to considering
only weak tachyon fields with $T^2 \ll T^3$, so
the exact form of the higher-order terms shall not
affect us.

\subsection{The black hole}

The black hole is a solution of the equations of motion \noeq{beteq}
with the tachyon field set to zero.  We shall consider these equations
in two convenient coordinate systems:
the conformal and the linear-dilaton gauges. The linear-dilaton gauge is
defined by taking the dilaton to be
proportional to a Liouville-like coordinate $\varphi$:
\beq
\Phi = -\sqrt2 \, \varphi \comma
\label{lindil}
\eeq
where we have inserted the factor of $-\sqrt2$ to match the background charge
of the dilaton in the $c=1$ string theory, $Q/2=-\sqrt{-c/12} = -\sqrt{2}$
\cite{DDK}.  We have also chosen an $x$-coordinate so that
$G_{x \varphi}=0$, used the freedom to redefine $x \to x'(x)$ to make $\pa_x$
the Killing vector of the metric, and normalized $x$ so that the solution
reduces to the flat background of the $c=1$ theory in the weak-coupling limit
$\varphi \to \infty$.  Then one finds a unique one-parameter family
of solutions to eqs\eq{beteq} with the tachyon field set to zero,
given by the linear dilaton of eq\eq{lindil} and the
metric \cite{Wadia} (see also section~\ref{back-sec}):
\beq
\d s^2 = \frac{1}{1-a \ep{-2}} \, \d \varphi^2 \pm
\left(1 - a \ep{-2} \right) \, \d x^2 \stop
\label{bhm}
\eeq
(Here the $+$ sign gives the Euclidean solution, and the $-$ sign the
Minkowski one.)   The parameter $a$ is the
ADM mass of the black hole \cite{Witten},
and one recovers
the usual linear dilaton solution
of the $c=1$ theory (without the cosmological constant) by setting
$a$ to zero.  The
mass can usually be changed by a constant shift of the
dilaton field, but here we have fixed this freedom by our definition of
the Liouville coordinate in eq\eq{lindil}.

We should stress that, as pointed out in
ref.~\cite{Witten}, the time coordinate in the black hole must be taken to
be $x$, and not $\phi$ as is usually done in the Liouville form of the
$c=1$ string.  This means that only the Euclidean version of
eq\eq{bhm} can be compared to the $c=1$ theory.
Although we shall often consider the Euclidean and Minkowski metrics of
eq\eq{bhm} together, and shall use the same notation and
nomenclature for the two cases, one should bear in mind that their
spacetimes are very different.  The Minkowski metric is
indeed that of a black hole, as can be seen by comparing the structure of
its zeroes and infinities to those of the standard
four-dimensional black hole in Schwarzschild coordinates: In terms of the
coordinate
\beq
z \equiv a \, \ep{-2} \comma
\label{defz}
\eeq
which we shall use extensively in this paper, the asymptotic
region of the black hole is given by $z \to 0$, the horizon is at
$z=1$, and the singularity is at $z=\infty$.  The ``Euclidean black
hole'' has the same asymptotic behaviour, but the horizon at $z=1$
becomes simply a coordinate singularity where the spacetime ends,
and the full space is given by the ``cigar'' of Witten \cite{Witten}.

These structures can be seen more clearly in the
``conformal gauge'', where the metric is taken to be proportional to the
Euclidean/Minkowski metric.  In the Euclidean case this coordinate system
is obtained by defining
\marginnote{Let u -> Sqrt(a)u, etc}
\beq
u \, \equiv\, \sqrt{\frac{(1-z)}{2z}} \, e^{i \sqrt2 x} \stop
\label{uvdef}
\eeq
Then the metric and the dilaton become \cite{Wadia}
\beq
\d s^2 = \frac{\d u \, \d \bar{u}}{1+2 u \bar{u}} \eqand
\label{bhm-con}
\eeq
\beq
\Phi = - \half \, \log a - \half \, \log \left({1+2 u \bar{u}}\right) \comma
\label{dil-con}
\eeq
so the free parameter in this gauge is a constant
shift of the dilaton.  Note that $z$ indeed ranges only from $0$ to $1$,
since it is now defined by
\beq
z = \frac{1}{1+2 u \bar{u}} \stop
\label{zdef}
\eeq
The asymptotic region is again $z \to 0$, but now the ``horizon'' at
$z=1$ is simply the origin of the complex $u$ plane.  This leads to an
important feature of the Euclidean black hole:  In order to avoid a
conical singularity at the origin, $z$ must be well defined.  From its
definition in eq\eq{uvdef} this means that
$\sqrt2 \, x$ is an angular coordinate, so
$x$ must be compactified on a circle of radius $R=1/\sqrt{2}$.
{}From the point of view of the asymptotic $c=1$ theory, this means that
$x$ is compactified with one-half of the self-dual radius!
The black hole thus constrains the radius of $x$, which
would be arbitrary in the Liouville theory.  A similar result was found
by Witten in the gauged WZW formulation of the black hole, where the radius
was found to be $\sqrt{k'/2}$ for large $k'$ (which is unfortunately
equal to $1/4$).  In our case, as well, the radius $1/2$
is also only valid
to lowest order in $\alpha'$.  This should be borne in mind when
comparing the black hole states to those of the $c=1$ theory.  The fact
that we are at $1/2$ the self-dual radius means that states in the
black hole can only have $x$-momenta\footnote{The winding states of the
$c=1$ theory can not be seen in the effective theory.} which are {\sl
integral} multiples of $\sqrt2$, and so describe only the integral-spin
$W_\infty$ states of Polyakov \cite{Kleb}.  However in the exact BRST
cohomologies of the gauged WZW theory \cite{Distler,Eguchi} and in the metric
of ref.~\cite{Dijkgraaf}, which is supposed to describe the theory to
all orders of $\alpha'$, the radius of $x$ is $R=\sqrt{2k}=3/\sqrt 2$,
or 3 times the black-hole value.  We shall return to the issue of the
exact metric in section \ref{exact}.

To get the Minkowski-space black hole, one Wick rotates $x \to i
x$.  Then $\bar{u}$
becomes independent of $u$, and should rather be called $v$, and $u$ and $v$
become null coordinates.
In this form, one can see that the Minkowski metric indeed has the Penrose
diagram of a black hole.  $z \to 0$ is again the asymptotic region of
the space.  $z=1$ now means $u v=0$, and so gives the horizon of the black
hole.  There is clearly no singularity of the metric or dilaton at the
horizon, so $z$ is no longer restricted to be $\le 1$.  In fact the
metric can be continued up to $z=\infty$, which is the singularity of
the black hole.  In the Minkowski case, there is no longer any
compactification of $x$, but the fact that it is compactified in the
Euclidean case implies a temperature of the Minkowski black hole of
$1/\sqrt2\pi$, or
$1/3\sqrt2\pi$ for the exact metric.

\section{The static tachyon in the black hole background}
\subsection{The tachyon and its boundary conditions}

In order to solve for the tachyon in the black-hole background,
it is useful to write its equations of
motion from eqs\eq{beteq} in terms of the coordinates $z$ and $x$.  The
equation then becomes
\beq
z(1-z) \, T'' - z \, T' + \frac{1}{4z} \, T \pm \frac{1}{8 z (1-z)} \, \ddot T
          = 0 \comma
\label{tachyoneq}
\eeq
with primes denoting derivatives with respect to $z$, and dots derivatives
with respect to $x$. Since this equation is linear and $x$ is a Killing
direction of the metric, one can Fourier expand the tachyon in terms of
tachyons with fixed $x$-momenta. We shall consider a general tachyon in
section \ref{moving}.  It is instructive to first consider the static
tachyon $T_{(0)}$ in detail. In the static case the Minkowski and
Euclidean equations are the same, and if one lets $T_{(0)} \to \sqrt{z} \,
F(z)$, they reduce to the hypergeometric equation:
\beq
z(1-z) \, F''(z) + (1-2z) \, F'(z) - \nfrac{1}{4} F(z) = 0 \stop
\label{tachyon}
\eeq
Thus
two linearly independent solutions
of the tachyon equation \noeq{tachyoneq}
are\footnote{In this case the two solutions
are equal to $2/\pi \sqrt z$ times the complete elliptic functions
$K(\sqrt z)$ and $K'(\sqrt z)$, respectively.} \cite{grin}:
\beq
\eqalign{
T_{(0)}^a (z) &= \sqrt{z} \, F(\half, \half, 1, z) \cr
T_{(0)}^b (z) &= \sqrt{z} \, F(\half, \half, 1, 1 - z) \comma
\label{tilde}
}
\eeq
with $F(\alpha,\beta,\gamma,z)$ is the standard ${}_2 F_1$ hypergeometric
function.

Asymptotically, the first solution $T_{(0)}^a (z)$ behaves as
\beq
T_{(0)}^a \onArrow{z \to 0} \sqrt{a} \, \ep{-} \, + \, \calo
(\ep{-3}) \stop
\eeq
This is the standard discrete zero-momentum tachyon which has been studied as
a perturbation of the black hole solution in \cite{Lykken}.
However, at the horizon it becomes
\beq
T_{(0)}^a \onArrow{z \to 1} -\, \frac{1}{\pi}\,
  \log \left( \frac{1-z}{16} \right) +
\hbox{vanishing terms,}
\label{tilde-ugh}
\eeq
and so blows up.  (It also has a cut from $1$ to $\infty$ in the
Minkowski black hole.)
Since $T$ is a scalar quantity, one would
expect that this divergence should invalidate the solution.
We shall indeed see
that this tachyon
induces a singularity in the metric when we study the back-reaction of the
tachyon on the black hole in section~\ref{back0-sec} and in the
appendix.
$T_{(0)}^a$ is therefore {\it not\/} a legitimate background for the tachyon in
the metric of the black hole.  Even if such a solution were meaningful
in the full theory, with higher order terms in the potential
stabilizing the tachyon,
it could not be studied in our approximation of neglecting higher powers
of the tachyon field.

We are thus left with the second solution $T_{(0)}^b (z)$.  More precisely,
we define
\beq
T_{(0)} = t_0 \, \sqrt{z} \, \F \comma
\label{thetach}
\eeq
where the constant $t_0$ should be small, since we are considering a weak
tachyon, and since we have solved for the tachyon in
the background of the unperturbed black hole.
This tachyon is a monotonically increasing function of $z$, and is
well behaved at the horizon, where it tends to $t_0$.
Asymptotically, it behaves as
\beq
T_{(0)} \onArrow{z \to 0} - \, \frac{\sqrt{a} \, t_0}{\pi} \, \ep{-} \,
    \left( \log \frac{a}{16} - \sqrt8 \, \varphi \right) \, + \,
    \calo (\varphi \, \ep{-3}) \semi
\eeq
it is thus a mixture of the usual zero-momentum tachyon and the
``cosmological constant'' operator \marginnote{nomenclature?} of the $c=1$
theory.  We thus see that
choosing a sensible boundary condition at the horizon leads
one to a combination of the two possible boundary conditions in the
asymptotic region of the black hole. We shall see that this behaviour is
generic, and occurs also for non-zero momenta tachyons and for the
discrete states.  In previous works it has been argued that the physical
tachyon is not $T$, but $S \equiv e^{-\Phi} T$ \cite{Lykken}; for our
tachyon, $S$ behaves asymptotically as $\varphi$, so it would not be
considered a valid solution.  However, looking
at the sigma-model equations of motion of eq\eq{beteq}, one sees that $T$ and
not $S$ is the physical quantity, and we shall indeed see that the back
reaction of $T_{(0)}$ is sensible.

In the Euclidean black hole, the spacetime stops at $z=1$.  In the
Minkowski case, one has to continue to the singularity of the black hole, at
$z\to\infty$.  For large $z$, $T_{(0)}$ behaves as
\beq
T_{(0)} \onArrow{z \to \infty} \frac{t_0}{\pi} \, \log (16 z) \comma
\label{sing}
\eeq
so our demand that the tachyon remain weak, so that $T^2 \ll T^3$,
restricts the validity of the tachyon solution to $z \ll \exp (\pi/t_0)/16$.
For $t_0$ small, this means that the tachyon solution is valid to near the
singularity of the black hole, although one can never actually reach the
singularity.

\subsection{The solution to the dilaton-graviton equations with tachyons}
\label{back-sec}

In order to find the back reaction of the tachyon
on the metric we need to
return to the full equations of motion of the theory, as given in
eqs\eq{beteq}.  Using the 2-dimensional identity
\beq
R_{\mu\nu} = \half\,g_{\mu\nu} R \comma
\eeq
the graviton and dilaton equations can be rewritten as the traceless equation
\beq
g_{\mu\nu} \nabla^2\Phi - 2\nabla_{\mu}\nabla_{\nu}\Phi = \half
g_{\mu\nu} (\nabla T)^2 -\nabla_{\mu}T\nabla_{\nu}T \comma
\label{fir}
\eeq
the dilaton equation
\beq
\nabla^2\Phi -2 (\nabla\Phi)^2 = -T^2 -4 \comma
\label{dilback}
\eeq
and the curvature scalar equation
\beq
R = 4(\nabla\Phi)^2 -8 -2 T^2 - (\nabla T)^2 \stop
\label{curv-s}
\eeq
In the linear-dilaton gauge, the dilaton is again defined by
$\Phi = -\sqrt2 \, \varphi $;
also noting the form of the black hole metric \noeq{bhm},
it is convenient to parameterize the metric by \cite{Lykken}
\beq
\d s^2 = \frac{1}{f} \, \d \varphi^2 \pm f \, h\, \d x^2 \stop
\label{bhdm}
\eeq
Then, the two independent components of eq\eq{fir} become
\beq
\eqalign{
\frac{h'}{h} &= -2z \, {T'}^2 +\frac1{4 z f^2 h}\, \tdot^2 \eqand \cr
\frac{\fdot}{f} &= 2 z \, \tdot T' \comma
\label{eqh}
}
\eeq
and eq\eq{dilback} becomes
\beq
f - z f' = 1 + \,\frac{T^2}{4}\, -z^2 f \, {T'}^2 +\,\frac1{8 f h} \, \tdot^2
\stop \label{eq2}
\eeq
The curvature scalar equation
\noeq{curv-s} becomes
\beq
R \, (f,h) = 8f-8 -2 \, T^2 - 8z^2(1-z) \, {T'}^2 \mp \frac{{\dot T}^2}{1-z}
    \comma
\label{eqr}
\eeq
but one does not need its explicit form, since it follows from the
Bianchi identities of the theory.
Similarly, consistency between the
equations for $\fdot$ and $f'$ imply the tachyon equation of motion
of eq\eq{beteq}.

If one sets the tachyon to zero, one immediately sees that $f_0=1-z$ and
$h_0=1$ are solutions to the equations of motion, giving us the black-hole
metric of eq\eq{bhm}. Since we are considering weak tachyon fields (having
solved for the tachyon in the background of the black hole without
matter), we can substitute $f \to f_0$ and $h \to h_0$ in the RHS of
eqs\eq{eqh} and \noeq{eq2}, and solve them perturbatively.  Thus
\beq
\eqalign{
h &= 1 + \int^z d z \left( -2 z \, {T'}^2  +\frac1{4z(1-z)^2}\,
          \tdot^2 \right) + A(x) \eqand \cr
f &= 1-z + z \int^z d z \left( -\frac1{4z^2}\, {T^2} +(1-z) \, {T'}^2
         -\,\frac1{8z^2(1-z)} \, \tdot^2 \right) + z B(x) \cr
     &= 1-z + 2 z (1-z) \int^x d x  \left( \tdot \, T'\right)  +C(z)    \stop
\label{pert}
}
\eeq
The arbitrary function $A(x)$ in $h$ can be fixed by demanding that $h\to1$
asymptotically.  Also, equating the two solutions for $f$ fixes $B(x)$ and
$C(z)$ up to
an arbitrariness of the form $f \to f + \alpha z$; this is a shift in the
mass $a$ of the original black hole, and has no physical meaning.

\subsection{The back reaction of the static tachyon }
\label{back0-sec}

The back reaction of the tachyon on the black hole is not linear, since
the tachyon appears quadratically in the graviton and dilaton equations of
motion.  It is thus necessary to calculate the back-reaction of a tachyon
which consists of a sum of fields with different momenta.  This shall be
done in section~\ref{back-ar-sec}.  Here we shall concentrate on the
back reaction of a static tachyon, in order
to understand the physics of the back reaction.
In this section we carry out the calculation in
the linear dilaton gauge; the calculation is
repeated in the conformal gauge in the appendix, where the
physical picture is somewhat clearer.
The
reader with a particular partiality to either gauge may concentrate on the
appropriate section.

Knowing the induced metric, we can now prove our statements
on the boundary conditions of the tachyon that we asserted in the previous
section.
In the static case the metric does not depend on $x$, which is
still a Killing direction, and the last equation in eq\eq{pert} carries no
information.  The other integrations in eqs\eq{pert} can be carried out
explicitly, using the tachyon equation of motion of eq\eq{tachyoneq}.  This
gives
\beq
\eqalign{
f &= (1-z) \left( 1+z \, T_{(0)} \, T_{(0)}' \, \right) \comma \cr
h &= 1 + c -\frac1{2z} \, T_{(0)}^2 +
     2(1-z) \, T_{(0)} \, T_{(0)}' -
           2 z (1-z) \, T_{(0)}^{\prime 2} \stop
\label{fh-res}
}
\eeq
Now, using eq\eq{eqr} for the curvature scalar, one sees that the
back reaction
of a tachyon containing the ``bad'' solution
$T_{(0)}^a$ of eq\eq{tilde} leads to a
curvature singularity of the form
\beq
R \sim \frac1{1-z}
\eeq
at the horizon.  Thus, as was expected from its divergent
behaviour in eq\eq{tilde-ugh}, $T_{(0)}^a$
is not a valid solution in the
black-hole background.

We thus see that one must indeed restrict oneself to our tachyon
of eq\eq{thetach} which is well behaved at the horizon.
In this case eqs\eq{fh-res} become:
\beq
\eqalign{
f  &= (1-z) \left\{ 1+ \, \frac{t_0^2 z}2 \, \F \right. \, \times \cr
   & \qquad \qquad \qquad \left.
            \left(\F - \frac12 \, \Fd \right) \right\} \comma \cr
h  &= 1 + \frac{2\, t_0^2}{\pi^2} -\frac{t_0^2}{2} \, z \, \Fsq -
	\frac{t_0^2}8 \, (1-z) \, \Fdsq   \stop
}
\eeq
One can now see that this resulting metric still has the structure of a
black hole\footnote{See also the appendix.}: Asymptotically
$f(0) =h(0)=1$, so one has the usual flat metric.
$h(z)$ is a monotonic function that slowly decreases from 1
to \hbox{$h(1)=1-t_0^2(1/2-2/\pi^2) > 0$}.  $f(z)$ increases briefly from 1
to a maximum, from which it decreases to vanish at $z=1$.  (The
arbitrariness in $f(z)$ was fixed so that the position of the zero of
$f(z)$ was not perturbed.)  This means that $z=1$ is still a horizon of
the metric.
In the Minkowski case, one must still go beyond the horizon:  For
$z>1$, $f(z)$ is negative and is never again zero, so it does not
produce any further structure.  This means that, contrary to the
claims in refs.~\cite{Minic,Rama}, the horizon is not split and the
metric does not have a Reissner-Nordstrom form.  Far beyond
the horizon one could worry about the new singularity of the metric
where $h(z)$ becomes zero, at $z \sim \exp(\pi^2/t_0^2)/16$.   However,
this is beyond the limit of our quadratic tachyon approximation (see
the discussion after eq\eq{sing}).  In any case the curvature scalar is
nonsingular there, and this is only a coordinate singularity.  We
conclude that only the original singularity $z\to\infty$ remains.

It thus appears that the perturbed metric differs very slightly from the
original black-hole metric, with the same asymptotic behaviour, horizon
and singularity structure.  However there is one crucial difference:
In the linear dilaton gauge the ADM mass $M$ of the black
hole is defined so that $f \onArrow{z \to 0} 1-M \ep{-2}$ asymptotically.
Recalling the definition of $z$ in eq\eq{defz}, one sees that
the original black hole with $f=1-z$ has mass $a$.  In
the perturbed black hole with
\beq
f(z) \onArrow{z \to 0} 1-z \left(1- \frac{t_0^2}{2\pi^2}
\left(\log^2\frac{z}{16} + 2\,\log\frac{z}{16} \right) \right) \comma
\eeq
one sees that the tachyon has shifted the mass of the black hole to minus
infinity!  While this would presumably be disastrous in four
dimensions, it appears to create no troubles in the dilaton-graviton
theory.

We conclude that a static tachyon can live in the back hole
background, thus giving hair to the black hole.

\section{Moving Tachyons}
\label{moving}

\subsection{The tachyon and its boundary conditions}

We now turn to a general tachyon with momentum $p$, so that
$T \sim T_{(p)}(z) \, \eipx$.
(The $\sqrt2$ is for convenience.)
The tachyon equation of motion \noeq{tachyoneq} now becomes
\beq
z(1-z) \, T_{(p)}'' - z \, T_{(p)}' + \frac{1}{4z} \, T_{(p)}
     \mp \frac{p^2}{4 z (1-z)} \, T_{(p)} = 0 \stop
\label{tachyonp}
\eeq
We have seen that the solutions are constrained by their behaviour at
the horizon ($z\to 1$).  In the Minkowski case (the positive sign
in eq\eq{tachyonp}), one can see that the two possible
behaviours of the tachyon there are $T_{(p)} \sim (1-z)^{\pm i p/2}$.  Such
tachyons correspond to the ``principal continuous series'' representation
of the \sltr{}
algebra in refs.~\cite{Dijkgraaf,Distler}.  However these
solutions oscillate
wildly around $z=1$ and are not continuous there, so they must be
discarded.  We immediately conclude that {\it only static tachyons can
exist in the Minkowski black hole.\/}

In the Euclidean case, the solutions of eq\eq{tachyonp} behave
as a linear combination of
$(1-z)^{\pm p/2}$.  This means that the generic solution blows up there,
and one obtains a unique solution by demanding that
the tachyon be well defined at the horizon.  Its form is found by
letting
$T_{(p)}(z) \to \, z^{(1+|p|)/2} \, (1-z)^{|p|/2} \, F(z)$,
transforming eq\eq{tachyonp} into the hypergeometric equation
\beq
z(1-z) \, F''(z) + (1+|p|) \, (1-2z) \, F'(z) - \nfrac{1}{4}\, (1+2|p|)^2
\, F(z) = 0 \stop
\eeq
The tachyon that is well behaved at the horizon is
then given by \cite{grin}
\beq
T_{(p)}(z) = t_p \; z^{\frac{1+|p|}{2}} \, (1-z)^{\frac{|p|}{2}} \,
     \Fp \stop \label{nst}
\eeq
Its asymptotic behaviour can be seen by
using the transformation property of the hypergeometric function:
\beq
\eqalign{
F(\alpha,\beta,\gamma,1-z) &=
      \, \frac{\Gamma(\gamma) \, \Gamma(\gamma - \alpha - \beta)}
      {\Gamma(\gamma-\alpha) \, \Gamma(\gamma - \beta)}
      \; F(\alpha,\beta,\alpha + \beta - \gamma  + 1,z)  \; +  \cr
& \; z^{\gamma - \alpha - \beta} \, (1-z)^{1-\gamma} \,
      \frac{\Gamma(\gamma) \, \Gamma(\alpha + \beta - \gamma)}
      {\Gamma(\alpha)\, \Gamma(\beta)}
      \, F(1 - \alpha, 1 - \beta, \gamma - \alpha - \beta +1,z)\comma
\label{prop}
}
\eeq
turning $T_{(p)}$ into
\beq
\eqalign{
T_{(p)}(z) &= \vphantom{\frac{\Gamma^2}{\Gamma^2}}
    -\, \frac{t_p}{\sin \pi |p|} \, z^{\frac{1+|p|}{2}} (1-z)^{\frac{|p|}{2}}
    \, F\n4\left(\half + |p|,\half + |p|,1 + |p|,z\right) \cr
       &\qquad \qquad + t_p \, \frac{\Gamma(1+|p|)\Gamma(|p|)}
                      {\Gamma^2\!\!\left(\half+|p|\right)}
            \, z^{\frac{1-|p|}{2}}(1-z)^{-\frac{|p|}{2}}
	    \, F\n4 \left(\half - |p|,\half - |p|,1 - |p|,z\right) \stop
\label{t-asym}
}
\eeq
One can then see that asymptotically the tachyon has the form:
\beq
T \onArrow{z \to 0} \, A \, \epa-{(1+|p|)} \, \eipx
       + B \, \epa-{(1-|p|)} \, \eipx \comma
\eeq
for some non-zero constants $A$ and $B$, and so
becomes a linear combination of the anti-Seiberg
and Seiberg tachyons of the $c=1$ theory.  This immediately leads us
to our most important result for tachyons with momentum: {\sl If $|p| >1$
the Seiberg component of the tachyon blows up asymptotically, so there is
no tachyon solution that is
well-behaved both asymptotically and at the horizon.\/}

The discrete
tachyons at $p=\pm1$ need to be studied separately; they are interesting
both because they are on the boundary of the allowed range of momenta, and
because if one indeed takes the radius of $x$ to be $1/\sqrt2$, as
suggested by the metric, only tachyons
with $p=0$ and $p=\pm1$ remain as viable solutions on the black hole.
The $p=\pm1$ tachyons are given by eq\eq{nst}:
\beq
T_{(\pm1)} = z \sqrt{1-z} \, F\n4\left(\thrhalf,\thrhalf,2,1-z \right) \stop
\eeq
Their asymptotic behaviour can not be found from eq\eq{t-asym},
which breaks down, but by examining its limit one
can see that they tend to $4/\pi$ asymptotically,
and so are well defined.

It is thus reasonable to conclude that only tachyons with $|p| \le 1$ can
exist in the Euclidean black hole; we shall immediately confirm this by
studying the
back reaction of a general tachyon.

\subsection{Back reaction}
\label{back-ar-sec}

In general, a moving tachyon breaks the black hole killing symmetry
and introduces an explicit dependence of
the metric \noeq{bhdm} on $x$.  The metric is still given by
eqs\eq{pert}, where $T$ is now a linear combination of the tachyons of
eq\eq{nst} with different momenta:
\beq
T = \sum_p \, T_{(p)} (z) \, \eipx \stop
\eeq
The integrations for $f$ can again be carried out explicitly, giving
\beq
f = (1-z) \left( 1
   + z \, \sum_{p} \, T_{(p)} \, T_{(-p)}' \,
   + z \sum_{p + q \neq 0} \, \frac{p \, T_{(p)} \, T_{(q)}' + q \, T_{(q)}'
   \, T_{(p)}} {p+q} \eix{(p+q)}
   \right) \stop
\eeq
$f$ clearly vanishes at the horizon, and asymptotically its $(p,q)$
term is proportionate to $T_{(p)} \, T_{(q)}$.
The $h$ integration in eq\eq{pert}
can not be done explicitly, but one can see that $h$
is finite at the horizon and has the same asymptotic
behaviour as $f$.
Using eq\eq{eqr},
one can also see that the curvature scalar $R$ is finite at the horizon, and
has the same asymptotic behaviour
as $f$.  Thus if the tachyon has components with momenta $|p| >1$, so that
$T_{(p)}$ blows up asymptotically, the
resulting metric is sick.  For tachyons
with momenta $|p| \le 1$, one sees that the modified metric is still
that of a black hole---it is asymptotically flat as $z \to 0$, and has
a horizon at $z=1$.

We conclude that tachyons with $|p| >1$ do not exist on the black hole. In
the Euclidean black hole (the 3) tachyons with momenta $|p| \le 1$ are good
solutions, perturbing the black hole and giving it ``hair''. The Minkowski
black hole supports only the static tachyon.

\section{Discrete states}
\label{discrete-sec}

\subsection{The BRST cohomology of the \sltrouo{} theory}

Having examined tachyons in the black-hole metric in some detail,
we would now like to
consider what happens to the other states in the black hole.
Since we are thinking of the black hole as the solution of a string
theory, one would expect to have an infinite number of massive states
arising from the mode-expansion of the string, and
since the theory is two-dimensional and is intimately
related to the $c=1$ string, one would expect these states to be discrete
states at fixed momenta.
Studying these states is not a trivial extension of the tachyonic
case, since the approach of coupling fields to the
sigma-model effective action of the string that we used there
has not been developed for massive states.  We shall thus have to
use a more indirect approach.

One might expect that the correct calculation to do would be to
actually find the states of the \sltrouo{} CFT underlying the black hole.
In fact, the complete analysis of the BRST cohomology of the
\sltrouo{} CFT has already been carried out.  This has been done
in two different
approaches:  First, the current algebra can be bosonized, using the Wakimoto
free-field representation of  the \sltr{} algebra \cite{Wakimoto}. The
bosonization of the $U(1)$ leads to an $x$ coordinate of radius
$R=\sqrt{2k}=3/\sqrt 2$, and the bosonization of the \sltr{} currents
lead to a Liouville-like field $\varphi$ with the appropriate background
charge. The
operators in the theory are then very similar to those of the $c=1$ theory
with radius $R=3/\sqrt 2$, and one can show that the spectra of the two
theories are identical
\cite{Eguchi}.  In the figure we show the states in one chiral sector of the
theory with the same ghost number as that of the tachyon.  The
other approach is to carry out the analysis directly in the current algebra
\cite{Distler}.  In that case $p_x$ and $p_\varphi$ are defined in terms
of the $J$
and $M$ of the \sltr{} algebra.  Here one finds that there are
extra states in addition to those of the bosonized theory;
these are also shown in the figure.

\setlength{\unitlength}{2cm}
\thicklines

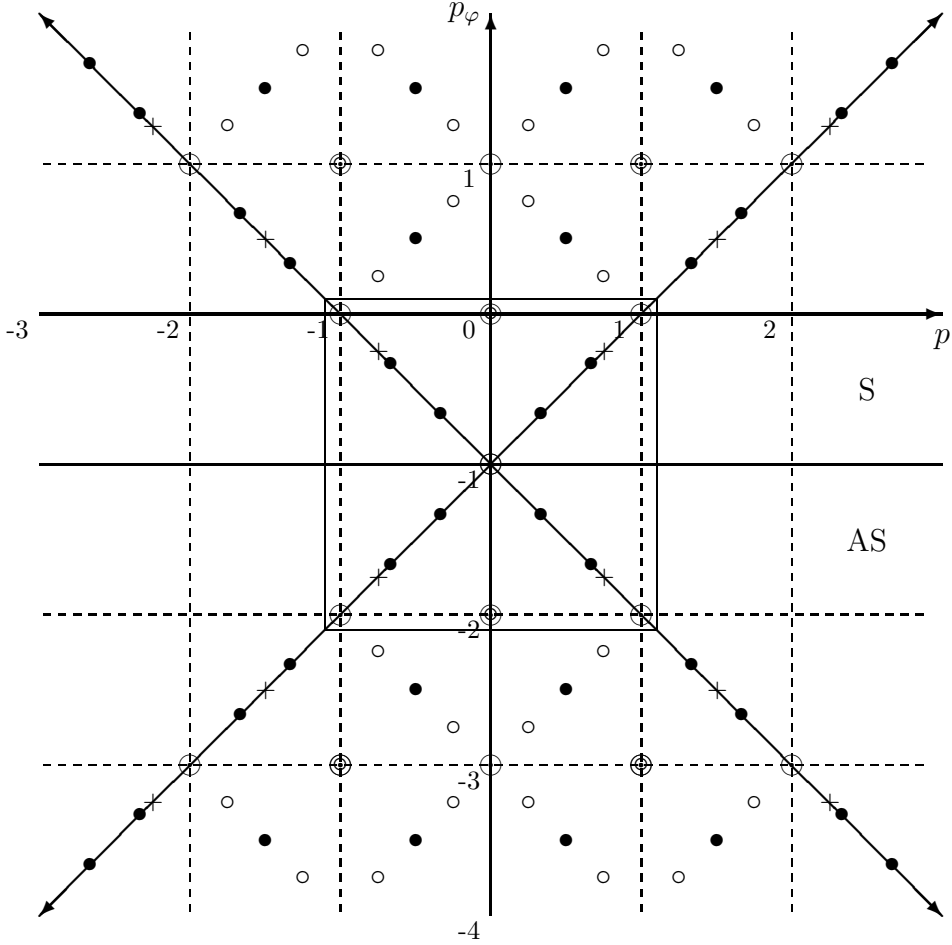
\begin{figure}
\begin{center}

\begin{picture}(6,7)(-3,-4.5)

\put(0,-4){\vector(0,1){6}}
\put(-3,0){\vector(1,0){6}}
\put(-3,-1){\line(1,0){6}}
\put(0,-1){\vector(1,1){3}}
\put(0,-1){\vector(1,-1){3}}
\put(0,-1){\vector(-1,1){3}}
\put(0,-1){\vector(-1,-1){3}}

\put(-.6,2){\makebox(.6,0)[r]{ $p_\varphi$ }}
\put(3,0){\makebox(0,-.1)[t]{$p$}}
\put(2.5,-1.5){\makebox(0,0){AS}}
\put(2.5,-.5){\makebox(0,0){S}}

\thinlines

\put(-1.1,-2.1){\framebox(2.2,2.2){}}

\multiput(-2.975,-3)(.1,0){59}{\line(1,0){.05}}
\multiput(-2.975,-2)(.1,0){59}{\line(1,0){.05}}
\multiput(-2.975,1)(.1,0){59}{\line(1,0){.05}}

\multiput(-2,-3.975)(0,.1){59}{\line(0,1){.05}}
\multiput(-1,-3.975)(0,.1){59}{\line(0,1){.05}}
\multiput(1,-3.975)(0,.1){59}{\line(0,1){.05}}
\multiput(2,-3.975)(0,.1){59}{\line(0,1){.05}}

\put(0,1){\makebox(-.3,-.2){ \footnotesize 1 }}
\put(0,-1){\makebox(-.3,-.2){ \footnotesize -1 }}
\put(0,-2){\makebox(-.3,-.2){ \footnotesize -2 }}
\put(0,-3){\makebox(-.3,-.2){ \footnotesize -3 }}
\put(0,-4){\makebox(-.3,-.2){ \footnotesize -4 }}

\put(-3,0){\makebox(-.3,-.2){ \footnotesize -3 }}
\put(-2,0){\makebox(-.3,-.2){ \footnotesize -2 }}
\put(-1,0){\makebox(-.3,-.2){ \footnotesize -1 }}
\put(0,0){\makebox(-.3,-.2){ \footnotesize 0 }}
\put(1,0){\makebox(-.3,-.2){ \footnotesize 1 }}
\put(2,0){\makebox(-.3,-.2){ \footnotesize 2 }}

\multiput(-2,-3)(1,1){5}{\makebox(0,0){ $\odot$ }}
\multiput(2,-3)(-1,1){5}{\makebox(0,0){ $\odot$ }}
\multiput(-2.6666667,-3.6666667)(1,1){6}{\makebox(0,0){ $\bullet$ }}
\multiput(2.6666667,-3.6666667)(-1,1){6}{\makebox(0,0){ $\bullet$ }}
\multiput(-2.3333333,-3.3333333)(1,1){6}{\makebox(0,0){ $\bullet$ }}
\multiput(2.3333333,-3.3333333)(-1,1){6}{\makebox(0,0){ $\bullet$ }}
\multiput(.75,-.25)(.75,.75){3}{\makebox(0,0){ \footnotesize + }}
\multiput(-.75,-.25)(-.75,.75){3}{\makebox(0,0){ \footnotesize + }}
\multiput(.75,-1.75)(.75,-.75){3}{\makebox(0,0){ \footnotesize + }}
\multiput(-.75,-1.75)(-.75,-.75){3}{\makebox(0,0){ \footnotesize +
}}

\multiput(-1,1)(1,0){3}{\makebox(0,0){ $\odot$ }}
\put(0,0){\makebox(0,0){ $\odot$ }}
\put(0,-2){\makebox(0,0){ $\odot$ }}
\multiput(-1,-3)(1,0){3}{\makebox(0,0){ $\odot$ }}

\multiput(-1,1)(2,0){2}{\makebox(0,0){ $\circ$ }}
\put(0,0){\makebox(0,0){ $\circ$ }}
\put(0,-2){\makebox(0,0){ $\circ$ }}
\multiput(-1,-3)(2,0){2}{\makebox(0,0){ $\circ$ }}

\multiput(-1,-3)(2,0){2}{\makebox(0,0){ $\odot$ }}
\multiput(-.5,.5)(1,0){2}{\makebox(0,0){ $\bullet$ }}
\multiput(-.5,-2.5)(1,0){2}{\makebox(0,0){ $\bullet$ }}
\multiput(-1.5,1.5)(1,0){4}{\makebox(0,0){ $\bullet$ }}
\multiput(-1.5,-3.5)(1,0){4}{\makebox(0,0){ $\bullet$ }}
\multiput(-1.75,1.25)(.5,.5){2}{\makebox(0,0){ $\circ$ }}
\multiput(-.75,.25)(.5,.5){4}{\makebox(0,0){ $\circ$ }}
\multiput(1.75,1.25)(-.5,.5){2}{\makebox(0,0){ $\circ$ }}
\multiput(.75,.25)(-.5,.5){4}{\makebox(0,0){ $\circ$ }}
\multiput(-1.75,-3.25)(.5,-.5){2}{\makebox(0,0){ $\circ$ }}
\multiput(-.75,-2.25)(.5,-.5){4}{\makebox(0,0){ $\circ$ }}
\multiput(1.75,-3.25)(-.5,-.5){2}{\makebox(0,0){ $\circ$ }}
\multiput(.75,-2.25)(-.5,-.5){4}{\makebox(0,0){ $\circ$ }}

\end{picture}
\end{center}

\caption{
The free-field and Kac-Moody chiral cohomology states of the \sltrouo{}
current algebra at $k=9/4$, split into Seiberg and anti-Seiberg states
by the line through $p_\varphi = -1$.  The discrete states common to
both cohomologies are denoted by large \hbox{circles $\odot$} for the integral
momenta states, and \hbox{``blobs'' $\bullet$} for states with half-integral
momenta.  The extra discrete states of the Kac-Moody cohomology are
depicted by small \hbox{circles $\circ$}.  Tachyons with no windings are
depicted by large circles for integral momenta, and blobs for momenta
in $\ZZ/3$, and the pure-winding tachyons are depicted by plus signs.
The box encloses those states for which neither their Seiberg and
their anti-Seiberg Liouville momenta are positive.
}

\label{plot}
\end{figure}

It would thus appear that we know which massive states appear in the
black hole.  However, there are many issues to settle.  The first is which (if
either) of the
two approaches discussed above is relevant to the black hole.  Here it may
be appropriate to recall another case where this issue has arisen---in
Polyakov's light-cone gauge approach to the $c=1$ string \cite{Pol}.  In
this gauge, the $c=1$ theory is written as a matter field $x$ coupled to a
(somewhat strange) \sltrouo{} theory.  It was found there that the
current-algebra analysis again leads to the presence of extra states in the
theory \cite{us};  since this theory should be equivalent to the Liouville
theory in the conformal gauge, this strongly suggests that the bosonized
approach is preferred.
It may also be worth noting that at the first level with $p=0$, where the
bosonized approach leads to one state and the Kac-Moody approach two, the
string
contains only the single ``discrete graviton state''
$\alpha_{-1}^\mu \tilde\alpha_{-1}^\nu$.

Even once one has decided which BRST cohomology is preferred, there are still
clearly several differences between the states of the black hole and the
spectrum of the \sltrouo{} CFT.  First, as we have already mentioned,
the black hole solution has been
derived only at the $O(\frac1{k^2})$ approximation to the CFT, while the
BRST analysis is exact.  One implication of this
is that the radius of $x$ in the black hole is $R=1/\sqrt2$, instead
of $R=3/\sqrt2$, so states in the black hole have only
``integral momenta'' $p=n$, instead of $p=n/3$.  One could, of course,
also expect other changes to the spectrum.
This problem will be rectified in
our discussion of the exact \sltrouo{} metric in
section~\ref{exact}. Another problem that is basic to all effective
space-time approaches to the string is that one can see only modes with
momenta, and not the intrinsically ``stringy'' winding states.
This is unavoidable, but
may be more of a technical problem then a deep issue in the theory.

A more interesting problem, and one which we shall hope to examine in our
space-time approach, is that the BRST cohomology analysis may simply
not be directly relevant to the physical spectrum of the
theory. In either approach the BRST analysis is carried out using
a free field representation of the \sltr{} current algebra which,
although useful for simplifying the analysis, obscures the
underlying geometry of the theory.  We have already seen that the BRST tachyons
do not appear directly in the spectrum of the black hole, and that
its tachyonic  spectrum is far smaller than that indicated
by the BRST analysis.
Another example of such a situation is again
the $c=1$ string; here the BRST analysis leads to a doubling of the
spectrum into Seiberg and anti-Seiberg states.  It is
now generally accepted that, while this may be true for the case of zero
cosmological constant $\mu$, if $\mu \neq 0$ the anti-Seiberg states are not in
the
spectrum of the $c=1$ theory, since they induce a catastrophic change to the
background of the theory.  For example, the ${\overline W}_{1,0}^- W_{1,0}^-$
\marginnote{Yaron?} operator changes the $c=1$ theory into the black hole!

\subsection{Discrete states in a nontrivial metric}

In the rest of this section, we shall thus attempt to analyse the discrete
states in a way that is related to our analysis of the tachyon.  The
most direct way of doing this would be
to find the physical states in the black hole
from the BRST cohomology in the black hole background: $Q \ket{\Psi} =
0$, $\ket{\Psi} \equiv \ket{\Psi} + Q \ket{\chi}$.  This is well
defined since any background of the string leads to a nilpotent
\marginnote{word?} BRST operator.  Equivalently, one could look at the
equations of motion and gauge invariance of the linearized string field
theory in this background.  However, neither of these methods
have been developed for string backgrounds that are more complicated
than small perturbations of flat space.  We shall thus use a poor-mans
approach, using the simplifying fact that,
asymptotically, the black hole reduces to the $c=1$ background of a
flat metric with a linear dilaton.

Let us thus briefly recall some facts
about the discrete states of the $c=1$ theory.
We shall be concerned only with the relative cohomology states with the
same ghost number as that of the tachyon, since only these can correspond to
physical states in the spacetime.  Since, as we have stated above, the
space-time approach can not deal with winding states, this leaves us
with only the Seiberg and anti-Seiberg states $W_\infty$ states:
\beq
{\overline W}_{J,M}^+ W_{J,M}^+ \qquad {\rm and} \qquad
{\overline W}_{J,M}^- W_{J,M}^- \comma
\label{W}
\eeq
where, \apriori, $J$ is a positive integer or half integer, and $|M| < J$.
However, if we define
states with behaviour $\epa{}{p_\varphi} \eipx$ to have
$x$-momentum $p$ and ``Liouville momentum'' $p_\varphi$, one
sees that the above states have $p=M$ and $p_\varphi = \pm J -1$.
Since we
are at one half of the self-dual radius in the black hole, or at
three halves the self-dual radius in the exact metric, $M$ and $J$ are
restricted to being integral in our case.
The states of eq\eq{W} can be built up using only $x$ oscillators.
Since these states are in the BRST cohomology, and since they do not
contain any ghost oscillators, they satisfy the Virasoro constraints
$\bar{L}_n =L_n =0$ for $n \ge 1$ and $\bar{L}_0 = L_0 = 1$.  On states
without winding, this last equation becomes
\beq
p^2+N_L - p_\varphi^2 - 2 p_\varphi = 1 \comma
\label{L0}
\eeq
where $N_L$ is the left-handed (say) level of the state measured from the
tachyon,
and its two solutions
give the Seiberg and anti-Seiberg Liouville momenta
\beq
p_\varphi^{S,AS} = -1 \pm \sqrt{p^2 + N_L}
\label{liouv}
\eeq
of the discrete state.

{}From a string field theory point of view, each of the discrete states
comes from a linear combination of many fields.
To give a concrete example, the discrete states at level 2
(which do not exist in our
case since $J=1/2$) are
\beq
{\overline W}_{\frac32,\frac12}^\pm W_{\frac32,\frac12}^\pm = \frac14
\left( \frac{i}{\sqrt{2}} {\bar{\pa}}^2 x + ({\bar{\pa}} x)^2 \, \right)
\left( \frac{i}{\sqrt{2}} \pa^2 x + (\pa x)^2 \, \right)
\, \epa{}{\left(\pm \frac32 -1\right)} \, \eix{\left(\frac12\right)}
\comma
\label{examp}
\eeq
and the analogous
${\overline W}_{\frac32,-\frac12}^\pm W_{\frac32,-\frac12}^\pm$ states with
$p = -1/2$.
These states come from expanding the
string field to level 2\footnote{Here, for simplicity,
we ignore terms containing $b_{-1} c_{-1}$ or $\tilde{b}_{-1} \tilde{c}_{-1}$,
since they do not appear in the discrete states.},
\beq
\Psi \sim A_{\mu\nu,\lambda\sigma} \, \alpha_{-1}^\mu \alpha_{-1}^\nu
         \tilde{\alpha}_{-1}^\lambda \tilde{\alpha}_{-1}^\sigma
      + A_{\mu\nu,\lambda} \, \alpha_{-1}^\mu \alpha_{-1}^\nu
         \tilde{\alpha}_{-2}^\lambda
      + A_{\lambda,\mu\nu} \, \alpha_{-2}^\lambda
         \tilde{\alpha}_{-1}^\mu \tilde{\alpha}_{-1}^\nu
      + A_{\mu,\nu} \, \alpha_{-2}^\mu \tilde{\alpha}_{-2}^\nu
\comma
\eeq
and using the (linearized) BRST condition $Q \Psi = 0$
and the equivalence $\Psi \equiv \Psi + Q \chi$.
This leave one with only the non-vanishing components
\beq
A_{xx,xx} \sim A_{xx,x} \sim A_{x,xx} \sim A_{x,x} \sim
\epa{}{\left(\pm \frac32 -1\right)} \, \eix{\left(\pm\frac12\right)}
\stop
\label{dsex}
\eeq

In principle, one would now like to carry out the same calculation
in the black-hole background.
Note that the $L_0=1$ equation of \noeq{L0} can be written covariantly
on each field $A_{{\mu}_1 \ldots {\mu}_j}$ of the string field as
\beq
    \nabla^2 A_{\mu_1 \ldots \mu_j}
    - 2\nabla\Phi \cdot \nabla A_{\mu_1 \ldots \mu_j}
   - m^2 A_{\mu_1 \ldots \mu_j} = 0 \comma
\label{aeq}
\eeq
where $m$ is the mass of the field, given by
\beq
N_L = 1 + \frac{m^2}{2} \stop
\eeq
It is thus reasonable to take eq\eq{aeq} to be the correct equation of the
theory
in a nontrivial background, although this is by no means proven, since
terms involving the Riemann tensor could also appear.  It
is encouraging, however, that no such terms have been found in the effective
sigma model couplings of the tachyon \cite{Tseytlin}.
Note that
the covariant derivatives in eq\eq{aeq} mix the different components of the
$A_{\mu_1 \ldots \mu_j}$'s if one has a non-trivial metric.
This means that the discrete states in the black hole
can not have only $x$-components nonvanishing,
and must also have nonvanishing $\varphi$- and mixed components, which
should be suppressed asymptotically.

One could also covariantize the $L_n=0$
equations, which should all be compatible, since the conformal
invariance of the background ensures the existence of a Virasoro
algebra.  However, since the resulting BRST analysis would be very difficult,
we shall rather think of
eq\eq{aeq} as being the
equation of motion of the theory in Siegel gauge, where all the fields
of the string are decoupled, and the kinetic operator is simply
$L_0-1$.  This equation of motion can be derived from the effective
sigma model action in this
gauge:
\beq
S = \int{\rm d}^2x \, e^{-2\Phi}\sqrt{G} \left(
  R - 4(\nabla\Phi)^2 + c/3  + {\sum_{A's}}^{\p2\prime}  \left\{ \left(
  \nabla A_{\mu_1 \ldots \mu_j} \right)^2 + m^2 A_{\mu_1 \ldots \mu_j}^2
     \right\} \right) \comma
\label{Siegel}
\eeq
where the sum is over all of the massive states of the string, as well
as the tachyon.  (The massless states are, of course, treated exactly
in the original sigma-model terms, and are not included in the sum.)

Being in Siegel gauge means that one has lost all information about the
gauge-invariance of the theory.  We shall restore this
information by making the \ansatz{} that the physical states of the theory
are found by solving the equation of motion of eq\eq{aeq}, and demanding
that the solutions tend asymptotically to the discrete states of the $c=1$
theory.  Equivalently, one can take the discrete states in the cohomology
of the asymptotic $c=1$ theory and continue them into the black hole
using eq\eq{aeq}.  It is reasonable to expect that this procedure does
indeed reproduce the spectrum of the black hole.  This is
supported by the fact that the free-field \sltrouo{} cohomology for the
black hole is, at least formally, in a one-to-one correspondence with
that of the $c=1$ theory, so one would not expect this procedure to
break down.  However, as in the tachyon case, we shall see that the
constraint that the resulting states be well-behaved over the entire
black-hole spacetime will drastically truncate the spectrum.

\subsection{The solution of the discrete states in the black hole}
\label{discrete-con}

To find discrete states in the target space of the black
hole, we now have to solve the equation of motion eq\eq{aeq} for the various
fields $A_{\mu_1 \ldots \mu_j}$ of the string field theory, and demand that
they asymptotically approach the $A_{x^j}$'s of the $c=1$ discrete states.
Since eq\eq{aeq} becomes a set of coupled second order differential equations
in the $(x,\varphi)$ or $(x,z)$ coordinate system,
we shall rather work in the conformal gauge defined by
eqs\eq{bhm-con} and \noeq{dil-con}.  Here, as is well known from using
complex coordinates on Riemann surfaces, the covariant derivatives do not
mix the $u$ and $\bar u$ components of fields, and indeed
the only nonvanishing Christoffel symbols are
\beq
\Gamma_{u u}^u = -\frac{2\bar{u}}{1+2 u \bar{u}} \comma
\label{offel}
\eeq
and its complex conjugate.  Eq\eq{aeq} then splits into separate equations
for each component of $A_{\mu_1 \ldots \mu_j}$, which we denote by $A_{u^n
\bar{u}^m}$, somewhat symbolically since the order of the indices is
important.  The complication of using these coordinates is that, in order to
match the asymptotic form of the discrete state to those of the $c=1$
theory, one has to return to the $(x,\varphi)$ coordinates.  Using
eq\eq{uvdef}, this transformation is given,  again somewhat symbolically, by
\beq
\eqalign{
A_{x^s \varphi^t} &= \sum \, i^{s(n-m)} \, z^{-(n+m)/2} \,
     (1-z)^{(s-t)/2} \, e^{i \sqrt{2} (n-m) x} \; A_{u^n \bar{u}^m} \cr
  &\onArrow{\varphi \to \infty} \sum \, i^{s(n-m)} \, a^{-(n+m)/2}
     \, \epa{}{(n+m)} \, e^{i \sqrt{2} (n-m) x} \; A_{u^n \bar{u}^m}
\stop
\label{trans}
}
\eeq

Eq\eq{aeq} is still a second order {\it partial\/} differential equation
for the $A_{u^n \bar{u}^m}(u,\bar{u})$'s,
but one can reduce it to an ordinary differential equation using the fact
that $x$ is a Killing direction of the metric.
We thus choose the fields $A_{u^n \bar{u}^m}$ to have
a well-defined $x$--momenta: In the Euclidean case
\beq
\eqalign{
A_{u^n \bar{u}^m} &\equiv A_{u^n \bar{u}^m}^{(p)} (z) \,
                \eix{(p-n+m)} \cr
            &= A_{u^n \bar{u}^m}^{(p)} \left(\frac1{1+2u \bar{u}}\right)
               \, \left(\frac{u}{\bar{u}}\right)^{(p-n+m)/2} \comma
\label{field-def}
}
\eeq
where the peculiar factor of $p-n+m$ in the momentum dependence
is chosen because, as is seen from their definition in eqs\eq{uvdef}, the
coordinate $u$ ``has momentum 1''.  With this convention, the
linear-dilaton-gauge fields $A_{x^s \varphi^t}$ of eq\eq{trans} will indeed
have momenta $p$.  In the last line of eq\eq{field-def}, we have used
the conformal-gauge
definitions of $z$ and $x$ from eqs\eq{uvdef} and \noeq{zdef}.
Now substituting the definition of eqs\eq{field-def} into eq\eq{aeq}, and using
the Christoffel symbols of eq\eq{offel}, one finally obtains
the hypergeometric-type equation:
\beq
\eqalign{
& z(1-z)\,\pa_z^2 \, A_{u^n \bar{u}^m}^{(p)} - (z +(n+m)(1-z)) \,
   \pa_z A_{u^n \bar{u}^m}^{(p)}
   - \frac{(p+m-n)^2}{4(1-z)} \, A_{u^n \bar{u}^m}^{(p)} \cr
&\qquad \qquad + \frac{1}{4z} \, ( (n+m+1)^2  - p^2-N_L) \,
        A_{u^n \bar{u}^m}^{(p)} -m n \, A_{u^n \bar{u}^m}^{(p)} = 0 \stop
\label{long}
}
\eeq

The Minkowski case is obtained by letting $p \to -i p$; then at the
horizon the solutions of eq\eq{long} tend to a linear combination
of $A_{u^n \bar{u}^m}^{(p)} \sim (1-z)^{\pm (i p-n+m)/2}$, and one can again
conclude that only static states can exist.

In the Euclidean case the solutions to eq\eq{long} have the form
$(1-z)^{\pm (p-n+m)/2}$ around the horizon.  One can not demand that $A_{u^n
\bar{u}^m}^{(p)}$ be finite since, being a tensor, it is not a physical
quantity.  However its ``norm'' $|A_{u^n \bar{u}^m}^{(p)}|^2 = (g^{u
\bar{u}})^{n+m} A_{u^n \bar{u}^m}^{(p)} A_{\bar{u}^n u^m }^{(p) *}$ is
physical, and must be finite.  Since $g^{u \bar{u}}=2/z$ is simply a
constant at the horizon, we conclude that $A_{u^n \bar{u}^m}^{(p)} \sim
|A_{u^n \bar{u}^m}^{(p)}|$ must
behave as $(1-z)^{|p-n+m|/2}$ at the horizon. Eq\eq{long} therefore has a
unique solution, which is given by:
\beq
A_{u^n \bar{u}^m}^{(p)} = z^{\alpha_+} \, (1-z)^{|p-n+m|/2} \,
           F \n2 \left(\beta_+,\beta_-,1+|p-n+m|,1-z \right) \comma
\eeq
with
\beq
\eqalign{
2 \alpha_\pm &= 1+n+m \pm \sqrt{p^2 + N_L} \eqand \cr
2 \beta_\pm &= 1 + |p-n+m|  \pm (n-m) + \sqrt{p^2 + N_L} \stop
}
\eeq
Using the transformation of eq\eq{prop},
one finds that $A_{u^n \bar{u}^m}^{(p)}$ has the
asymptotic behaviour
\beq
A_{u^n \bar{u}^m}^{(p)}  \onArrow{z \to 0} A \, z^{\alpha_+} +
           B \, z^{\alpha_-} \stop
\eeq
The interpretation of this behaviour is not obvious, since
the metric $g^{u \bar{u}}=2/z$ diverges asymptotically.
However, since the physical quantity $|A_{u^n \bar{u}^m}|$ behaves as:
\beq
|A_{u^n \bar{u}^m}|  \onArrow{z \to 0}
       C \, z^{(1+\sqrt{p^2 + N_L} )/2} +
       D \, z^{(1-\sqrt{p^2 + N_L} )/2} \comma
\label{dis-as}
\eeq
one sees that one has a well-behaved discrete state in the black hole
only if $p^2+N_L \le 1$.  Since the level $N_L$ is a non-negative
integer, this means that either $N_L = 0$, $|p| \le 1$ or $N_L =1$, $p =0$.

The physical meaning of this constraint can be seen by returning to the
asymptotic discrete state $A_{x^{n+m}}$ defined via eq\eq{trans}.  This has
the form
\beq
\eqalign{
A_{x^{n+m}}  &\onArrow{z \to 0}
                E \, z^{(1+\sqrt{p^2 + N_L} )/2} \, \eipx +
              F \, z^{(1-\sqrt{p^2 + N_L} )/2} \, \eipx \cr
       &= G \, \epa{}{p_\varphi^{AS}} \, \eipx
       + H \, \epa{}{p_\varphi^{S}} \, \eipx \comma
\label{phy-as}
}
\eeq
where in the last equation we have used the Liouville momenta
$p_\varphi^{S,AS}$ of eq\eq{liouv}. This means that the discrete state
$A_{x^{n+m}}$ is again a combination of the anti-Seiberg and Seiberg
discrete states of the $c=1$ theory, and the condition $p^2+N_L \le 1$ is
simply the condition that neither the Seiberg nor the anti-Seiberg Liouville
momenta is positive.  Examining the figure, one can see that the only
states satisfying this condition are the discrete tachyons (with $N_L = 0$
and $|p| \le 1$) that we have already found, and the massless ``discrete
graviton'' state(s), with $N_L=1$ and $p=0$.  However, in our approach to
the string-field states we do not add extra linearized gravitons and
dilatons to the action, since all the massless fields are already included
in the fully nonlinear sigma-model action around which we perturb.
``Dilaton-graviton hair'' with $N_L=1$ simply corresponds to the number
of free
parameters in the black-hole solution, and we have seen that the only
freedom in the 2D black hole solution is a constant shift of the dilaton,
corresponding to the only parameter, the ADM mass $a$.

We concluded that {\it the black hole has no W--hair!\/}

\section{Tachyon and W--hair in the exact \sltrouo{} background }
\label{exact}

\subsection{The exact \sltrouo{} metric and dilaton}

The effective space-time black hole background of eqs\eq{lindil} and
\noeq{bhm} corresponds
to the semiclassical $O(\frac{1}{k'^2})$ approximation of the
\sltrouo{} gauged
WZW conformal field theory\footnote{This can be easily seen, for
instance, by comparing central charges of the theories\cite{Witten}.}.
(Here $k'=k-2$, where $k$ is the central charge
of the \sltr{} Kac-Moody algebra.)  An exact effective
space time background {\it for states without windings\/} was proposed
by Dijkgraaf \etal{} \cite{Dijkgraaf} for arbitrary $k$.  It is derived
by expressing the $L_0, \bar{L}_0$
Virasoro generators of the \sltrouo{} gauged WZW theory as
differential operators on the \sltr{} group manifold, and
identifying the operator $L_0 + \bar{L}_0$ with the target
space Laplacian of the sigma model.

After rescaling the $r$ and $x$ coordinates to asymptotically approach the
the $c=1$ theory, the thus derived metric and dilaton take the Euclidean and
Minkowski forms:
\beq
\eqalign{
\d s^2 &= \d r^2 \pm \beta^2(r)\, \d x^2 \eqand \cr
\Phi &= \Phi_0 -\frac12 \,  \log\left(\frac{\sinh \left(\sqrt{2/k'} \, r
       \right )}{\beta(r)} \right) \comma
\label{exactmet}
}
\eeq
with $\beta(r)$ given by
\beq
\beta(r) = \left( \frac{k}{k' }\coth^2 \left(\frac{r}{\sqrt{2k'}}\right)
      - \frac{2}{k'}    \right)^{-\frac{1}{2}} \stop
\label{betr}
\eeq
It has been verified up to three loops in \cite{Tseytlin} and up to four
loops in \cite{Liverpool} that the dilaton-graviton background of
eq\eq{exactmet} is indeed conformally invariant.  For large $k$, this
background reduces to the dilaton-graviton black hole.
For the background to give a 2-dimensional solution to the string, one needs
to take $k=9/4$, which we shall do from now on.

It is again useful to define a
coordinate $z$, in this case by
\beq
z \equiv \sech ^{2} \sqrt2 r \stop
\label{z}
\eeq
This transforms the metric and dilaton of eqs\eq{exactmet} to
\beq
\eqalign{
\Phi &= \frac12 \, \log \left(\frac{9 z}{a} \right)
             - \frac14 \, \log \left({1+8z} \right) \eqand \cr
\d s^2 &= \frac{1}{8z^2(1-z)} \, \d z^2 \, \pm \, \frac{1-z}{1+8z} \, \d x^2
       \comma
\label{metz}
}
\eeq
where we have written the constant shift of the dilaton in terms of the ADM
mass $a$.  These are of the same form as the black hole solution of
eqs\eq{lindil} and \noeq{bhm}, except for the factors of $1+8z$ (and of 9).
It is clear that
the metric has a horizon at $z=1$, corresponding to $r=0$, and
that the asymptotic region of the metric at $r\to \infty$ maps to $z=0$.
In the Euclidean case, one is again restricted to $0 \le z \le 1$.  In the
Minkowski case the metric has a different behaviour from that of the black
hole metric as $z\to\infty$: Since $g_{xx}$ no longer diverges, the metric
has only a coordinate singularity there.  It was argued in ref.~\cite{Teo}
that the spacetime could therefore be maximally extended by
linking together infinitely many copies of the metric through $z=\infty$.
However it is important to note that
the dilaton blows up as $z\to\infty$.  Since the dilaton is an
integral part of the 2-dimensional gravity theory, we shall regard
$z\to\infty$ as a singularity of the spacetime, and shall not allow
anything to pass through this point.  We shall also see that
matter fields such as the tachyon blow up as $z\to\infty$.

Since we are provided with a conjectured exact background of the
string, and since it has a structure very similar to that of
the black hole,
it is natural to ask
whether the conclusions of the previous sections concerning the tachyon
and W--hair remain valid also in this exact background.
In the rest of this section we shall
carry out the required analysis in order to answer
this question.  In the exact theory one can not study the
back-reaction of the fields explicitly, since the dilaton-graviton
action is an infinite series in $\alpha'$.  However, it is reasonable to
argue that any field with a finite well-behaved stress tensor has a
valid back-reaction on the metric and dilaton, and thus leads to a sensible
perturbation of the theory.

We should, perhaps, remind the reader that
``stringy states'' with nontrivial
windings can not be described in our
target space approach.  In principle, one should be able to describe states
with winding, but no momenta, using the dual metric of
ref.~\cite{Dijkgraaf}.  This is simply given by the region of
eqs\eq{metz} between $z=-1/8$ and $z=0$, and describes a space with a naked
singularity (at $z=-1/8$).
However it is known that there are difficulties
with describing the string by this dual metric.  For example, as was noted
in ref.~\cite{Dijkgraaf}, one can not see the
periodicity of $x$ from the target space metric.  From our point of view we
would not find any restrictions on the fields in this dual space, since there
is no horizon to fix their boundary conditions.

\subsection{Static and non-static \sltrouo{} tachyons}

If one assumes that the effective action of the tachyon still has
the form
\beq
S = \int{\rm d}^2x \, e^{-2\Phi}\sqrt{G} \left( \, (\nabla T)^2 -2 T^2 \right)
\comma
\eeq
the tachyon equation of motion is now given by
\beq
z(1-z) \, T'' - z  \, T' \pm \frac{1+8z}{8 z (1-z)}  \, \ddot T \pm
\frac{1}{4z} \, T = 0 \comma
\label{exacTz}
\eeq
in agreement with the equation derived from the gauged
WZW theory \cite{Dijkgraaf}.
Note, however, that the issue of higher order terms in the tachyon potential
still exists for the exact metric.
If one now considers a tachyon with momentum $p$, $T \sim T_{(p)}(z)
\, \eipx$, the equation of motion becomes
\beq
z(1-z) \, T_{(p)}'' - z  \, T_{(p)}' - \left( \frac{(\pm p^2 - 1)}{4z} \pm
\frac{9p^2}{4(1-z)} \right) T_{(p)} = 0 \stop
\label{exacTp}
\eeq
In the Minkowski metric, the solutions of this equation behave as
$T_{(p)} \sim (1-z)^{\pm 3 i p /2}$ around the horizon,
so one is again restricted to having
only static tachyons.  In the Euclidean case,
the solutions behave as
$T_{(p)} \sim (1-z)^{\pm 3|p|/2}$, so demanding good behaviour at the horizon
leads one to the unique well-behaved solution:
\beq
T_{(p)} (z) = t_p \, z^{\frac{1+|p|}{2}} \, (1-z)^{\frac{3|p|}{2}} \,
F\n4 \left(\nfrac12 + 2 |p|\,, \nfrac{1}{2} + 2 |p|\,, 1 + 3|p|\,, 1-z \right)
    \stop  \label{tpsol}
\eeq
This has a form very similar to that of the black hole tachyon of
eq\eq{nst}, and one is led to
similar conclusions for the exact tachyon.  Thus,
using eq\eq{prop}, one again sees that $T_{(p)}$ has
both anti-Seiberg and Seiberg components asymptotically,
\beq
T \onArrow{z \to 0} \, A \, \epa-{(1+|p|)} \, \eipx +
         B \, \epa-{(1-|p|)} \eipx \comma
\eeq
so demanding finite asymptotic behaviour again restricts one to
tachyons with $|p| \le 1$.  In the exact solution
$x$ has radius $3/\sqrt{2}$, as is also
seen directly from the gauged
WZW theory, so momenta are restricted to be multiples
of $1/3$.  We thus conclude that in the background of the exact solution
to the Euclidean \sltrouo{} theory,
the only tachyons that are well behaved are $T_{(0)}$,
$T_{(\pm1/3)}$, $T_{(\pm2/3)}$ and $T_{(\pm1)}$, and that only the static
tachyon can exist in the Minkowski solution.

\subsection{\sltrouo{} W--hair}

In section~\ref{discrete-sec} we showed that
W--hair does not exist for the black hole.  We argued there that the problem
could be reduced to solving the Siegel-gauge equation of
motion (neglecting possible heigher terms in the potential)
\beq
\nabla^2 A_{\mu_1 \ldots \mu_j}
    - 2\nabla\Phi \cdot \nabla A_{\mu_1 \ldots \mu_j}
    - m^2A_{\mu_1 \ldots \mu_j} = 0
\label{exaeq}
\eeq
for a general massive field
$A_{{\mu}_1 \ldots {\mu}_j}$ with mass $m$, and matching
the solution to the discrete states of the $c=1$ theory.
Here we would like to repeat this procedure in the exact dilaton-graviton
background of eqs\eq{exactmet}.  As in the black-hole case, a difficulty with
this procedure is that in the $(x,z)$ coordinates the various components of
$A_{{\mu}_1 \ldots {\mu}_j}$ mix.  This could be solved, as before, by solving
eq\eq{exaeq} in
conformal coordinates but, while this could certainly be carried out, it is
somewhat messy.  Since our final conclusions are based
only on the fact that the asymptotic fields have both
anti-Seiberg and Seiberg
components, we shall simplify our lives by
restricting ourselves to examining only
scalar pieces of the $A_{{\mu}_1 \ldots {\mu}_j}$'s, made by contracting them
with $G^{\mu\nu}$'s and $\varepsilon^{\mu\nu}$'.  This will be sufficient to
show that the discrete states can not exist in the exact
\sltrouo{} background.

We thus define, generically,
\beq
A\equiv G^{{\mu}_1{\mu}_2} \ldots \varepsilon^{{\mu}_i{\mu}_{i+1}} \ldots
        A_{{\mu}_1 \ldots {\mu}_j} \stop
\eeq
Since $A$ is a scalar, eq\eq{exaeq} reduces to
\beq
z(1-z) \, A_{(p)}'' - z  \, A_{(p)}' - \left( \frac{1}{4z} \,
 \left(\pm p^2 + \frac{m^2}{2} \right) \pm  \frac{9p^2}{4(1-z)} \right)
        A_{(p)} = 0 \comma
\label{extraeq}
\eeq
for an $A_{(p)}$ with the usual momentum dependence
$A \sim A_{(p)}(z) \eipx$.  Around the horizon, the Minkowski-space
solutions to this equation again behave as
$A_{(p)} \sim (1-z)^{\pm 3 i p /2}$, so only static discrete states can
exist.  In the Euclidean case the solutions behave as
$A_{(p)} \sim (1-z)^{\pm 3|p|/2}$
around the horizon.
Since $A_{(p)}$ is a scalar quantity it is physical, and must be regular
everywhere.  Therefore, there is a unique well-behaved solution for $A_{(p)}$,
which is given by:
\beq
A_{(p)} (z) = a_p \, z^{\frac{1+\sqrt{p^2 + N_L}}{2}} \,
       (1-z)^{\frac{3|p|}{2}}
      \, F\n4 \left(\beta,\beta , 1+ 3|p|,1 - z \right)
   \comma
\label{exasol}
\eeq
where $\beta$ is now given by
\beq
\beta = \frac12 \left( 1+3|p| + \sqrt{p^2 + N_L} \right) \stop
\eeq
As in section~\ref{discrete-sec}, we have defined the level
$N_L$ by $N_L = 1 + \frac{m^2}{2}$.
As expected, this solution exhibits a combination of
anti-Seiberg and Seiberg behaviour asymptotically:
\beq
\eqalign{
A &\onArrow{z \to 0} \,
C \, z^{\frac{1+\sqrt{p^2 + N_L}}{2}} \,  \eipx +
        D \, z^{\frac{1-\sqrt{p^2 + N_L}}{2}} \, \eipx \cr
  &= E \, \epa{}{p_\varphi^{AS}} \, \eipx
       + F \, \epa{}{p_\varphi^{S}} \, \eipx \comma
\label{exAasy}
}
\eeq
and so blows up except when $N_L = 0$ and $|p| \le 1$, or when $N_L =1$ and
$p =0$.

We can now reach our conclusion that one can not have discrete states in the
exact solution.   Since the potential discrete states are sums of
terms of the form $A_{x^j}$ asymptotically, the ``$A$'s'' obtained by
contracting the $A_{{\mu}_1 \ldots {\mu}_j}$'s with $G^{\mu\nu}$'s
are necessarily
nonvanishing.  We have shown that such $A$'s can only exist for
$N_L = 0$ and $|p|
\le 1$, giving the discrete tachyons of the previous section, and for
$N_L =1$ and $p =0$. This last case exists only for the ``discrete
graviton'' which, again is not taken as a linearized perturbation, since the
full nonlinear dilaton-graviton system has been solved.

We conclude that W--hair does not exist on the exact
\sltrouo{} target space, and that the black hole is completely described by
its ADM mass, and the 7
discrete tachyons with $|p| \le 1$.

\section{Conclusions}

In this paper we have studied the tachyon and $W_\infty$ states in the 2D
dilaton-graviton black hole of Witten, and in the exact metric proposed by
Dijkgraaf \etal{} to describe the \sltrouo{} conformal field theory.
We have used the effective sigma-model approach for the tachyon, and
have developed a mixed Siegel-gauge string field theory/sigma model
approach for the massive states.
We again warn the reader that
one can not deal with winding states from these space-time points of
view.

After solving their linearized equations of motion exactly, we
have seen that choosing states to be well behaved at the horizon of
the black hole forces them to be in a combination of Seiberg and
anti-Seiberg states asymptotically.  For these states to be well
behaved asymptotically, they then need to satisfy the condition that
neither the Seiberg nor the anti-Seiberg
Liouville momenta can be positive.  We thus disagree with
previous works on the subject, which all worked only with anti-Seiberg states
that blow up on the horizon.
Our condition means that there are no
$W_\infty$ states in the black hole, and that only the tachyons with
$|p| \le 1$ are good states.
Because the radius of $x$ in the black
hole is $1/\sqrt2$, one has
only the three tachyons $T_{(0)}$ and
$T_{(\pm 1)}$.  In the exact metric the radius is $3/\sqrt2$, so one
has the four additional states $T_{(\pm1/3)}$ and $T_{(\pm2/3)}$.
In the Minkowski black holes, only the static
tachyon survives.

The black hole is stable to the
back-reaction of these tachyons, so they can be regarded
as being hair of the black hole.
It is very interesting that this spectrum is so much
sparser than
the free-field BRST cohomology of the underlying \sltrouo{} conformal
field theory, which contains an infinite number of tachyon and
$W_\infty$ states.  We regard this as being
similar to the truncation of the spectrum of the $c=1$ theory to the
Seiberg sector, once the cosmological constant is turned on and the
Liouville geometry of the theory becomes relevant.
As far as the black hole is concerned, this has the implication that
the black hole is uniquely
described by its mass and the various tachyonic perturbations.  This
small amount of hair is clearly useless for maintaining
quantum coherence during any black-hole evaporation.
{}From the point of view of the underlying \sltrouo{} conformal field
theory, it is interesting to speculate that the specific tachyon
perturbations that we have picked out
may be related to integrable perturbations of the
theory.

\vskip 1 cm

{\large \bf \noindent Acknowledgments}

We are grateful to D.~Gross for a question to S.~Yankielowicz, leading
us to clarify the differences between the Euclidean and Minkowski black
holes.

\newpage

\appendix{The back reaction of the static tachyon in the conformal gauge}

In this section we derive the back reaction of the tachyon on the
black hole in the Minkowski version of the conformal gauge, in which
the global structure
of the metric is most easily seen.  The metric in these coordinates
is defined to be
\beq
\d s^2 = e^\sigma \d u \, \d v \stop
\label{met-c}
\eeq
In the following we shall need the Christoffel symbols and the Ricci tensor,
the only non-vanishing components of which are
\beq
\eqalign{
\Gamma_{u u}^u &= \pa_u \sigma \comma \cr
\Gamma_{v v}^v &= \pa_v \sigma \eqand \cr
R_{u v} &= \pa_u\pa_v\sigma \stop
\label{nonvan}
}
\eeq
As we saw in eqs\eq{bhm-con} and \noeq{dil-con}, the dilaton-graviton black
hole
solution in these coordinates is given by
\marginnote{Let u -> Sqrt(a)u, etc}
\beq
\eqalign{
\sigma &= - \, \log ( 1+2u v ) \comma \cr
\Phi &= - \half \, \log a - \half \, \log ( 1+2u v ) \comma
\label{bh}
}
\eeq
with the horizon and singularity located at $u v=0$ and $u v=-1/2$,
respectively, and the asymptotic region at $u v \to \infty$.
Since
the dilaton and the components of black hole metric depend only on the
product $u v$, the system has a Killing symmetry.  In these coordinates
$z$ is defined by
\beq
z = \frac{1}{1+2 u v} \comma
\eeq
and, as before, the asymptotic region is at $z=0$, the
horizon at $z=1$ and the singularity at $z \to \infty$.

We now turn to finding the backreaction of a tachyon on the metric in this
gauge.  For simplicity, we shall consider only a static tachyon (although
the general case could be considered as in section~\ref{discrete-con}); then
since $T_{(0)}$ depends only on $z$, $\sigma$ and $\Phi$ also continue to do so
after the backreaction.  The graviton $\beta$--function equation in
\noeq{beteq} gives rise to the two equations:
\beq
 \pa_z \biggl ( z(1-z) \pa_z (\sigma  -2 \Phi) \biggr )
     + z(1-z)  \,  T_{(0)}^{\prime 2} = 0 \eqand
\label{grava}
\eeq
\beq
\Phi'' + \frac{2}{z} \, \Phi' - \Phi' \sigma'
      -\half \, T_{(0)}^{\prime 2} =0  \comma
\label{gravb}
\eeq
and the dilaton  $\beta$--function equation in \noeq{beteq} becomes
\beq
\pa_z \left( z(1-z) \pa_z (\sigma -4 \Phi) \right) +
          z(1-z)(\, 4 \, {\Phi'}^2 + T_{(0)}^{\prime 2} \, )
          - \frac{1} {z^2} \left( 1 + \frac1{4z} \, T_{(0)}^2 \right) \,
          e^\sigma = 0 \stop
\label{dilb}
\eeq
The prime denotes differentiation with respect to $z$.
Before turning on the tachyon field, one can see that the general solution of
these equations\footnote{We show this in the static case, but it should be
true in general, as seen in the linear-dilaton gauge in \cite{Lykken}.  (See
also eqs\eq{eqh} and \noeq{eq2}.)}
is equivalent to the black hole of \noeq{bh}:
First, eq\eq{grava} can be integrated, giving
$\sigma-2\Phi = \gamma+\alpha \log(z/(1-z))$.
This ``$\alpha$--term'', which implies a coordinate divergence at the horizon,
can be eliminated by the conformal coordinate transformation $u' =
u^{1-\alpha}$, $v' = v^{1-\alpha}$. Now eqs\eq{grava},
\noeq{gravb} and \noeq{dilb} can be combined to give a first-order
differential equation for $\sigma$, which can be integrated.  This gives the
black hole of \noeq{bh}.

Now, to find the back reaction of the tachyon, first note that the
$z$ defined here is the same as that of the
linear-dilaton gauge (see the transformations of eq\eq{uvdef}),
so the tachyon is again given by eq\eq{thetach}:
\beq
T_{(0)} = t_0 \, \sqrt{z} \, \F \stop
\eeq
The back reaction of the tachyon on the black hole is found by
expanding eqs\eq{grava}, \noeq{gravb} and \noeq{dilb} around the black hole,
\ie{} $\sigma \to \log(z) + \tilde\sigma$ and
$\Phi\rightarrow \frac12\log(z/a) + \tilde \Phi$, and substituting for the
tachyon.  This gives the equations:
\beq
\eqalign{
& \pa_z \left( z(1-z) \pa_z (\tilde\sigma -2 \tilde\Phi) \right)
      + z(1-z) \, T_{(0)}^{\prime 2} = 0 \comma \cr
&\tilde\Phi'' - \frac{1}{2z}(\tilde\sigma' - 2\tilde\Phi')
              -  \half \, T_{(0)}^{\prime 2}  = 0 \comma \cr
&z (1-z)(\tilde\sigma''-4 \tilde\Phi'' ) + (1 - 2z) \, \tilde\sigma' + 4 z \,
\tilde\Phi' - \frac{1}{z}\tilde \sigma
+ z(1-z) \, T_{(0)}^{\prime 2} - \nfrac14 \, T_{(0)}^2 = 0\stop
\label{pert-con}
}
\eeq
The first equation of eq\eq{pert-con} can again be solved by integration.
Using  the tachyon equation of motion eq\eq{tachyoneq}, one finds
\beq
\sigma-2\Phi = \log(a) + \frac{t_0^2}{2} - \frac1{2z} \, T_{(0)}^2
           +(1-z)\, T_{(0)} \,T_{(0)}' \comma
\eeq
up to an ``$\alpha$--term'' which can be set to zero, as before, and a
constant which we fix by demanding that $\sigma-2\Phi=\log a$ at $z=1$.
Note that, unlike that unperturbed black hole, one can no longer have
$\sigma = 2(\Phi-\Phi_0)$ everywhere when tachyons are present.
This is in
contrast to studies of ``2D gravity'', where the black hole is coupled to
conformal matter that does not see the dilaton field, and one can choose the
$\sigma \equiv 2\Phi$ as a gauge \cite{Verlins}. Here
the tachyon couples to the
dilaton, as expected from the effective action of the string, making this
impossible.

The conformal factor $\sigma$ is once again found by combining the remaining
equations to get a first-order differential equation, which can be
integrated.  Remarkably, this integration can be done explicitly,
giving
\beq
\sigma = \log\left(z \right)+ 2 \left(\frac{t_0}{\pi} \right)^2 z -
  \frac1{2z} \, T_{(0)}^2 + 2 \, (1 - z) \, T_{(0)} \, T_{(0)}' -
   2 z \, (1 - z)^2 \, T_{(0)}^{\prime 2}  \stop
\label{sig-p}
\eeq
(The constant of integration, corresponding to a shift of the black-hole
mass, is fixed by demanding that $\sigma$ have no term linear in $z$
for large $z$.)  The resulting dilaton field is given by
\beq
\Phi = \frac12 \log\left(\frac{z}{a} \right)+ \left(\frac{t_0}{\pi} \right)^2 z
    -\frac{t_0^2}4 +  \frac12 (1 - z) \, T_{(0)} T_{(0)}' -
    z (1 - z)^2 \, T_{(0)}^{\prime 2} \stop
\label{phi-p}
\eeq

One can now easily see the geometric structure of this spacetime.
First, examining the form of $\sigma$ and $\Phi$, one sees that there are
no special points other than $z=0$, $1$ and $\infty$.  As $z\to 0$, the scalar
curvature\footnote{The reader who wishes to compare
this result with the linear-dilaton gauge result of eqs\eq{eqr} and
\noeq{fh-res} should be warned that after the tachyon perturbation the
``$z$'s'' of the two gauges are no longer identical!}  from eq\eq{nonvan}
\beq
R = -8 z -2 \left(\frac{t_0}{\pi} \right)^2 (1-z) + \frac1{4z} \, T_{(0)}^2
  - z(1 - z)(1 - 2 z) T_{(0)}^{\prime 2}
\eeq
becomes
\beq
R \onArrow{z \to 0} -8z \left( 1 - \left(\frac{t_0}{\pi} \right)^2
        \left(3+\log \frac{z}{16}\right) \right) \comma
\label{curv}
\eeq
so the spacetime is asymptotically flat, with a linear dilaton field.
$z=1$ is the horizon of the black hole, since it corresponds to $u v = 0$.
Unlike the linear-dilaton gauge, where this is a coordinate singularity, both
$\sigma$ and $\Phi$ are finite there, so one sees that there is no singularity
at the horizon.  There are then no singularities until $z\to\infty$, the
original singularity of the black hole.

We conclude that while the (static)
tachyon perturbation changes
the conformal factor and the form of the dilation,
the global structure of the black hole is unchanged.  The static tachyon
is thus sensible hair for the black hole.

\end{document}